\pdfoutput=1
\documentclass[12pt]{article}
\usepackage{graphicx}
\usepackage{subcaption}
\usepackage{amssymb}
\usepackage{amsmath}
\usepackage{amsfonts}
\usepackage{multirow}

\usepackage{xcolor}
\usepackage{url}
\usepackage{ulem}
\usepackage{float}
\usepackage{float}

\usepackage{pifont}
\newcommand{\cmark}{\ding{51}}%
\newcommand{\xmark}{\ding{55}}%

\usepackage{soul}
\usepackage[english]{babel}
\usepackage[T1]{fontenc}
\usepackage{lmodern}
\usepackage[utf8]{inputenc}  
\usepackage{bbm}

\usepackage{hyperref}
\hypersetup{
  colorlinks   = true, 
  urlcolor     = blue, 
  linkcolor    = black, 
  citecolor   = black 
}

\usepackage[numbers,sort&compress]{natbib}
\graphicspath{{plots/}}

\newcommand{\lsim}
{\;\raisebox{-.3em}{$\stackrel{\displaystyle <}{\sim}$}\;}
\newcommand{\gsim}
{\;\raisebox{-.3em}{$\stackrel{\displaystyle >}{\sim}$}\;}

\newcommand\tb{\tan\beta}

\newcommand\SB{s_\beta}
\newcommand\CB{c_\beta}

\newcommand\ra{\rightarrow}
\newcommand\tenp[1]{\times 10^{#1}}

\newcommand\ReDiag{\mathop{%
  \raise .5pt\hbox{[}%
  \widetilde{\mathrm{Re}}%
  \raise .5pt\hbox{]}}}
\newcommand\ReOffDiag{\mathop{%
  \raise .5pt\hbox{$\llbracket$}%
  \widetilde{\mathrm{Re}}%
  \raise .5pt\hbox{$\rrbracket$}}}

\newcommand\cL{{\cal L}}

\newcommand\SW{s_\mathrm{w}}
\newcommand\CW{c_\mathrm{w}}
\newcommand\MW{M_W}
\newcommand\MZ{M_Z}
\newcommand\Mh{M_h}

\newcommand\MA{M_A}

\newcommand\Sn{\tilde\nu}
\newcommand\Sl{\tilde l}

\newcommand\Slpm{\tilde l^\pm}

\newcommand\Sel[1]{\tilde e_{#1}}
\newcommand\Smu[1]{\tilde \mu_{#1}}

\newcommand\mse[1]{m_{\Sel{#1}}}
\newcommand\msl[1]{m_{\Sl_{#1}}}

\newcommand\Stau[1]{{\tilde\tau_{#1}}}
\newcommand\stau{\tilde \tau}

\newcommand\mL{m_{\tilde l_L}}
\newcommand\mR{m_{\tilde l_R}}
\newcommand\msnu{m_{\tilde \nu_l}}

\newcommand\ino[1]{\tilde\chi_{#1}}

\newcommand\chapm[1]{\ino{#1}^\pm}
\newcommand\champ[1]{\ino{#1}^\mp}
\newcommand\chap[1]{\ino{#1}^+}
\newcommand\cham[1]{\ino{#1}^-}
\newcommand\cha{\chapm}
\newcommand\mcha[1]{m_{\chapm{#1}}}
\newcommand\mchap[1]{m_{\ino{#1}^+}}

\newcommand\neu[1]{\ino{#1}^0}
\newcommand\mneu[1]{m_{\neu{#1}}}

\newcommand\refeq[1]{Eq.~(\ref{#1})}
\newcommand\refeqs[1]{Eqs.~(\ref{#1})}
\newcommand\refta[1]{Tab.~\ref{#1}}
\newcommand\refse[1]{Sect.~\ref{#1}}
\newcommand\refses[1]{Sects.~\ref{#1}}
\newcommand\citere[1]{Ref.~\cite{#1}}

\newcommand{\CP}{{\cal CP}}
\newcommand{\cp}{{\CP}}

\newcommand{\tev}{\,\, \mathrm{TeV}}
\newcommand{\gev}{\,\, \mathrm{GeV}}
\newcommand{\mev}{\,\, \mathrm{MeV}}

\newcommand\MO{\texttt{MicrOMEGAs}}
\newcommand\CM{\texttt{CheckMATE}}

\newcommand\fb{\ensuremath{\mbox{fb}}}
\newcommand\ab{\ensuremath{\mbox{ab}}}

\newcommand\ifb{\ensuremath{\fb^{-1}}}
\newcommand\iab{\ensuremath{\ab^{-1}}}

\newcommand\msele[1]{m_{\tilde{e}_{#1}}}
\newcommand\msmu[1]{m_{\tilde{\mu}_{#1}}}
\newcommand\mstau[1]{m_{\tilde{\tau}_{#1}}}

\newcommand{\br}{\text{BR}}

\newcommand{\sig}{\sigma}

\def\order#1{\ensuremath{{\cal O}(#1)}}
\def\reffi#1{\mbox{Fig.~\ref{#1}}}
\def\reffis#1{\mbox{Figs.~\ref{#1}}}

\def\ga{\gamma}
\def\De{\Delta}

\def\gmin2{\ensuremath{(g-2)_\mu}}
\def\amu{\ensuremath{a_\mu}}
\def \met  {\mbox{${E\!\!\!\!/_T}$}}
\newcommand{\ssi}{\ensuremath{\sig_p^{\rm SI}}}
\newcommand{\ssd}{\ensuremath{\sig_p^{\rm SD}}}
\newcommand{\Och}{\Omega_\chi h^2}

\definecolor{Orange}{named}{orange}
\definecolor{Purple}{named}{purple}
\definecolor{Lightblue}{cmyk}{0.9,0.1,0.1,0.3}
\definecolor{dgelborange}{cmyk}{0.,0.3,0.5, 0.}
\definecolor{Lila}{rgb}{0.5,0.,1}
\definecolor{Darkgreen}{rgb}{0.,.7,0.2}

\evensidemargin \oddsidemargin
\renewcommand{\arraystretch}{1.2}

\allowdisplaybreaks
\begin{document}
\thispagestyle{empty}

\def\thefootnote{\fnsymbol{footnote}}

\begin{flushright}
\mbox{}
IFT--UAM/CSIC--20-077\\
IPMU20-0057\\	
arXiv:2006.15157 [hep-ph]
\end{flushright}

\vspace{0.5cm}

\begin{center}

{\large\sc 
{\bf Improved \boldmath{\gmin2} Measurements and Supersymmetry
}}

\vspace{1cm}

{\sc
Manimala Chakraborti$^{1}$%
\footnote{email: mani.chakraborti@gmail.com}%
, Sven Heinemeyer$^{1,2,3}$%
\footnote{email: Sven.Heinemeyer@cern.ch}%
~and Ipsita Saha$^{4}$%
\footnote{email: ipsita.saha@ipmu.jp}%
}

\vspace*{.7cm}

{\sl
$^1$Instituto de F\'isica Te\'orica (UAM/CSIC), 
Universidad Aut\'onoma de Madrid, \\ 
Cantoblanco, 28049, Madrid, Spain

\vspace*{0.1cm}

$^2$Campus of International Excellence UAM+CSIC, 
Cantoblanco, 28049, Madrid, Spain 

\vspace*{0.1cm}

$^3$Instituto de F\'isica de Cantabria (CSIC-UC), 
39005, Santander, Spain
\vspace*{0.1cm}

$^4$Kavli IPMU (WPI), UTIAS, University of Tokyo, Kashiwa, Chiba 277-8583, Japan
}

\end{center}

\vspace*{0.1cm}

\begin{abstract}
\noindent
The electroweak (EW) sector of the Minimal Supersymmetric Standard Model
(MSSM) can account for variety of experimental data. The lighest
supersymmetric particle (LSP), which we take as the lightest neutralino,
$\neu1$, can account for the observed Dark Matter (DM) 
content of the universe via coannihilation with the next-to-LSP
(NLSP), while being in agreement with negative results from
Direct Detection (DD) experiments. Owing to relatively small production
cross-sections a comparably light EW sector of the MSSM is also in
agreement with 
the unsuccessful searches at the LHC. Most importantly, the EW sector of the
MSSM can account for the persistent $3-4\,\sig$ discrepancy between the
experimental result for the anomalous magnetic moment of the muon, \gmin2, and
its Standard Model (SM) prediction. Under the assumption that the $\neu1$
provides the full DM relic abundance we first analyze which mass ranges of
neutralinos, charginos and scalar leptons are in agreement with all
experimental data, including relevant LHC searches.
We find an upper limit of $\sim 600 \gev$ for
the LSP and NLSP masses. 
In a second step we assume that the new result of the 
Run~1 of the ``MUON G-2'' collaboration at Fermilab yields a precision
comparable to the existing experimental result with the same central
value. We analyze the potential impact of the combination of the Run~1
data with the existing \gmin2\ data on the allowed
MSSM parameter space. We find that in this case the upper limits
on the LSP and NLSP masses are substantially reduced by roughly $100 \gev$.
This would yield improved upper limits on these
masses of $\sim 500 \gev$. 
In this way, a clear target could be set
for future LHC EW searches, as well as for future high-energy
$e^+e^-$~colliders, such as the ILC or CLIC. 
\end{abstract}


\def\thefootnote{\arabic{footnote}}
\setcounter{page}{0}
\setcounter{footnote}{0}

\newpage


\section{Introduction}
\label{sec:intro}

One of the most important tasks at the LHC is to search for physics beyond the 
Standard Model (SM). This includes the production 
and measurement of the properties of Cold Dark Matter (CDM). 
These two (related) tasks will be among the top priority in the future
program of  high-energy particle physics. 

The high-energy searches are complemented by low-energy experiments that
search either for rare beyond the SM (BSM) decays, or for small
deviation of known SM processes from their SM prediction. Concerning the
latter the anomalous magnetic moment of the
muon, \gmin2\, plays a prominent role. The experimental result deviates
from the SM prediction by
$3-4\sig$~\cite{Keshavarzi:2019abf,Davier:2019can}. 
Improved experimental results are expected in the course of 2020 by the 
publication of the Run~1 data of the ``MUON G-2''
experiment~\cite{Grange:2015fou}.  

Another clear sign for BSM physics is the precise measurement of the CDM relic
abundance~\cite{Planck}. A final set of related constraints comes from
CDM Direct Detection (DD) experiments. The LUX~\cite{LUX},
PandaX-II~\cite{PANDAX} and XENON1T~\cite{XENON} experiments provide
stringent limits on the spin-independent DM scattering cross-section, \ssi.

\smallskip
Among the BSM theories under consideration the Minimal Supersymmetric
Standard Model  
(MSSM)~\cite{Ni1984,Ba1988,HaK85,GuH86} is one of the leading candidates.
Supersymmetry (SUSY) predicts two scalar partners for all SM fermions as well
as fermionic partners to all SM bosons. 
Contrary to the case of the SM, in the MSSM two Higgs doublets are required.
This results in five physical Higgs bosons instead of the single Higgs
boson in the SM.  These are the light and heavy $\cp$-even Higgs bosons, 
$h$ and $H$, the $\cp$-odd Higgs boson, $A$, and the charged Higgs bosons,
$H^\pm$.
The neutral SUSY partners of the (neutral) Higgs and electroweak gauge
bosons gives rise to the four neutralinos, $\neu{1,2,3,4}$.  The corresponding
charged SUSY partners are the charginos, $\cha{1,2}$.
The SUSY partners of the SM leptons and quarks are the scalar leptons
and quarks (sleptons, squarks), respectively.

The electroweak (EW) sector of the MSSM, consisting of charginos,
neutralinos and scalar leptons can account for a variety of experimental
data. Concerning the CDM relic abundance, the MSSM
offers a natural candidate, the Lightest Supersymmetric Particle 
(LSP), the lightest neutralino,~$\neu1$~\cite{Go1983,ElHaNaOlSr1984},
while being in agreement with negative results from
DD experiments.
On the other hand, the unsuccessful searches at the LHC can be attributed
to the rather small production cross-sections,
keeping a relatively light EW sector of the MSSM well alive.
Most importantly, the EW sector of the
MSSM can account for the persistent $3-4\,\sig$ discrepancy of \gmin2.

Various articles have investigated (part of) this interplay.
The impact of LHC Run~I searches in particular on the chargino/neutralino
sector of the MSSM have been discussed, among others, in
\cite{Bharucha:2013epa,Han:2013kza,Choudhury:2016lku}, and including
Run~II prospects in \cite{Datta:2016ypd,Chakraborti:2017vxz}.
A combination of Run~I and (then) current \gmin2\ data can be found
in \cite{Hagiwara:2017lse,Yanagida:2020jzy} while the  \gmin2\ without the LHC constraints were
discussed in \cite{Yin:2016shg,Yanagida:2016kag}.
Compressed chargino/neutralino spectra that are difficult to access at
the LHC in the context of DM bounds were discussed
in \cite{Chakraborti:2017dpu}. 
The direct searches at the LHC Run~I, the (then) current \gmin2\
deviation from its SM prediction, the measurement of the CDM relic
abundance and the limits from CDM DD experiments have been analyzed in a
global fit to the phenomenological MSSM with 11 parameter
(pMSSM11~\cite{AbdusSalam:2011fc}) in~\cite{Bagnaschi:2015eha}. It was
found that this model ``easily'' satisfies all the constraints together.
LHC Run~II data, the (then) current bound from \gmin2\ as well as DM
constraints were analyzed in several benchmark planes
in \cite{Datta:2018lup,Cox:2018qyi,Cox:2018vsv} and in two benchmark scenarios
w.r.t.\ the DM 
relic abundance in \cite{Abdughani:2019wai}. In the latter the
relevant LHC searches have been applied without a proper re-casting.
All Run~II data, again without proper re-casting, but with DM
data, was taken into account in \cite{Endo:2020mqz}, favoring models
with relatively heavy sleptons. 
Run~II data for chargino/neutralino searches
with some DM implications has also been presented
in \cite{Pozzo:2018anw,Athron:2018vxy}.

The aim of the paper is two-fold. Under the assumption that the $\neu1$
provides the full DM relic abundance we first analyze which mass ranges of
neutralinos, charginos and sleptons are in agreement with all
relevant experimental data.
Concerning the LHC searches we include all relevant existing data,
mostly relying on re-casting via \CM~\cite{Drees:2013wra,
Kim:2015wza, Dercks:2016npn}.  
In a second step we assume that the new result of the
Run~1 of the ``MUON G-2'' collaboration at Fermilab yields a precision
comparable to the existing experimental result with the same central
value. We analyze the potential impact of the combination of the Run~1
data with the existing result on the allowed
MSSM parameter space. The results will be discussed in the context of
the upcoming searches for EW particles at the HL-LHC. We will also comment
on the discovery prospects for these particles at possible future
$e^+e^-$~colliders, such as the ILC~\cite{ILC-TDR,LCreport} or
CLIC~\cite{CLIC,LCreport}.


\section {The electroweak sector of the MSSM}
\label{sec:model}

Here we briefly review the EW sector of the MSSM, consisting of charginos,
neutralinos and scalar leptons. The scalar quark sector is assumed to
be heavy and not to play a relevant role in our analysis. Throughout this
paper we also assume that all parameters are real, i.e.\ the absence of
$\CP$-violation.

MSSM neutralinos are the linear superpositions of the
neutral $SU(2)_L$ and $U(1)_Y$ gauginos and neutral higgsinos
$\tilde B, \tilde {W^3}$, $\tilde H_{u}^{0}$ and $\tilde H_{d}^{0}$,
respectively.
Their masses and mixings are determined by $U(1)_Y$ and $SU(2)_L$
gaugino masses $M_1$ and $M_2$, the Higgs mixing parameter $\mu$
and $\tb$, the ratio of the two
vacuum expectation values (vevs) of the two Higgs doublets of MSSM,
$\tb = v_2/v_1$.
The neutralino mass matrix in the basis
$(-i \tilde B, -i \tilde W^3, \tilde H_{u}^{0}, \tilde H_{d}^{0} )$
is given by
\begin{equation}
M_{N}=\left(
\begin{array}{cccc}
  {M_1} & 0 & -\MZ \CB \SW & \MZ \SB \SW \\ 0 &   {M_2} &
\MZ \CB \CW   & -\MZ \SB \CW  \\ -\MZ \CB
\SW & \MZ \CB \CW  & 0 &   {-\mu} \\ \MZ \SB
\SW &
-\MZ \SB \CW  &   {-\mu} & 0
\end{array} \right)
\end{equation}
where $\CB (\SB)$ denotes $\cos\beta(\sin\beta)$
and $\CW = \sqrt{1 - \SW^2} = \MW/\MZ$ denotes effective weak
leptonic mixing angle, with $\MW (\MZ)$ being the mass of the
$W$~($Z$)~boson. After diagonalization, the four eigenvalues
of the matrix give the four neutralino masses
$\mneu1 < \mneu2 < \mneu3 <\mneu4$.
As discussed above, the lightest neutralino, $\neu1$ is the LSP and
is assumed to yield the full CDM relic density, see \refse{sec:relic}
below.

The chargino mass eigenstates result as a mixing between the charged winos and
higgsinos $(\tilde {W^\pm}, \tilde H_{u/d}^{\pm})$ respectively with their mass
matrix given by,
\begin{equation}
M_{C}=\left(
\begin{array}{cc}
 {M_2} & \sqrt 2 \MW\CB \\\sqrt 2 \MW\SB &  {\mu }
\end{array} \right)
\end{equation}
Diagonalizing $M_C$ with a bi-unitary transformation, two chargino-mass
eigenvalues $\mcha1 < \mcha2$ can be obtained.

For the sleptons, as will be discussed below, we choose common soft
SUSY-breaking parameters for all three generations.
The charged slepton mass matrix can be written as,
\begin{equation}
{M_{\tilde L}}^2=\left(
\begin{array}{cc}
m_l^2 + m_{LL}^2 & m_l X_l\\
m_l X_l & m_l^2 + m_{RR}^2 \end{array} \right)
\end{equation}
 where
\begin{align}
m_{LL}^2 &=  \mL^2 + (I^{3L}_l - Q_l \SW^2)\MZ^2 \cos(2\beta)~, \nonumber\\
m_{RR}^2 &= \mR^2 + Q_l \SW^2 \MZ^2 \cos(2\beta)~, \nonumber\\
X_l &= A_l - \mu (\tb)^{2 I^{3L}_l}~. \nonumber\\
\notag
\end{align}
Here $l = e, \mu, \tau$ and $\mL$ and $\mR$ are the left- and
right-slepton 
mass input parameters, $I^{3L}_l$ and $Q_l$ denote the weak isospin and
electric charge of the lepton $l = e, \mu, \tau$.
We take the trilinear coupling
$A_l$ to be zero for all the three generations of leptons. Thus the
off-diagonal term $m_l X_l$ is given by $m_l \mu \tb$, and hence the mixing 
is significant only for the third generation.
Thus, for the first two generations, the mass eigenvalues can be approximated
as $\msl1 \simeq m_{LL}, \msl2 \simeq m_{RR}$.
In general we follow the convention that $\Sl_1$ ($\Sl_2$) has the
large ``left-handed'' (``right-handed'') component.
Besides the symbols equal for all three generations, we also
explicitly use the scalar electron, muon and tau masses,
$\mse{1,2}$, $\msmu{1,2}$ and $\mstau{1,2}$.

The sneutrino and slepton masses are connected by the usual
MSSM mass relation :
\begin{equation}
\msl1^2 = \msnu^2 - \MW^2 \cos(2\beta).
\end{equation}
Thus, overall the EW sector at the tree level
can be described with the help of six parameters: $M_1,M_2,\mu, \tb, \mL, \mR$.
Throughout our analysis we neglect $\cp$-violation and
assume $\mu, M_1, M_2 > 0 $.
From general considerations, even sticking to real parameters, one
could choose some (or all) of the mass parameters negative. However, it
should be noted that the results for physical observable are affected
only by certain combinations of signs (or phases in general). It is
possible, for instance, to rotate the phase of $M_2$ away, i.e.\ choose
(real and) $M_2 > 0$. This leaves in principle the signs of $M_1$ and
$\mu$ free. However, one of the main constraints we will take into
account is the anomalous magnetic moment of the muon, $\amu := \gmin2/2$,
see \refse{sec:gmin2}. Having $M_1$, $M_2$ and $\mu$ positive yields in
general a positive contribution to $\amu$, as it is required by
experimental data (see \refse{sec:gmin2}). Changing one or both signs of
$M_1$ and $\mu$ negative yields a substantially reduced or even negative
contribution to $\amu$. Consequently, having this constraint in mind,
our choice of only positive values is justified (see, however, the
discussion in \refse{sec:conclusion}).


\subsection*{The other MSSM sectors}

Following the stronger experimental limits from the
LHC~\cite{ATLAS-SUSY,CMS-SUSY},
we assume that the colored sector of the MSSM is sufficiently heavier
than the EW sector, and does not play a role in this analysis. For the
Higgs-boson sector we assume that the radiative corrections to the light
$\cp$-even Higgs boson (largely originating from the top/stop sector)
yield a value in agreement with the experimental data, $\Mh \sim 125 \gev$.
This naturally yields stop masses in the TeV
range~\cite{Bagnaschi:2017tru}, in agreement 
with the above assumption. Concerning the Higgs-boson mass scale, as
given by the $\cp$-odd Higgs-boson mass, $\MA$, we employ the existing
experimental bounds from the LHC. In the combination with other data,
this results in a non-relevant impact of the heavy Higgs bosons on our
analysis, as will be discussed below.


\section {Relevant constraints}
\label{sec:constraints}

In this section we briefly review the experimental constraints that are
relevant for the EW sector of the MSSM. They consist of direct searches
at the LHC, the \gmin2\ deviation from its SM prediction, the
measurement of the CDM relic abundance and the limits from CDM DD
experiments. A review about the combination of these effects in SUSY
(after LHC Run~1) can be found in~\cite{Bagnaschi:2015eha}. 


\subsection{Constraints from the LHC}
\label{sec:collider}

In the absence of color-sector SUSY particles within the LHC reach,
the production of electroweak gauginos ($\chap1 \cham1, \chapm1 \neu2$) and
sleptons ($\Sl_{L,R} \Sl_{L,R}$) are the most important search channels
at the LHC.
The ATLAS and CMS collaborations have searched for these processes
in a variety of final states. 
The LHC searches are usually interpreted in terms of ``simplified'' models
with specific assumptions on the compositions and branching ratios of
the SUSY particles. For example, all the searches described through
\refeq{3ldecviaslep}-(\ref{chaviawh}) assume $\chapm1 / \neu2$ and
$\neu1$ to be purely wino- and bino-like respectively 
so that $\mneu2 = \mcha1$.
However, the sensitivity of the searches may vary significantly
with the variation of the gaugino-composition as well as mass hierarchy
among the SUSY particles.

\begin{figure}[htb!]
        \centering
        \begin{subfigure}[b]{0.48\linewidth}
        \centering\includegraphics[width=\textwidth]{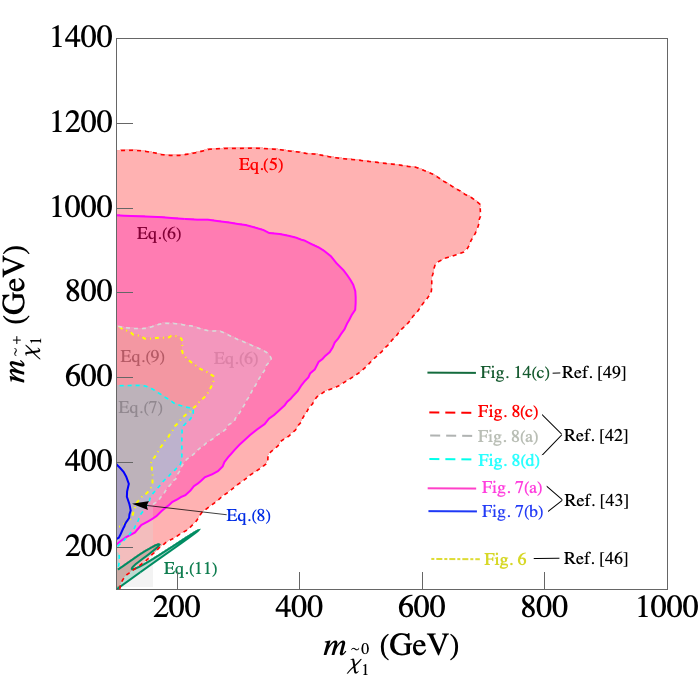}
                \caption{}
                \label{contour_mn1_mch}
        \end{subfigure}
        ~
        \begin{subfigure}[b]{0.48\linewidth}
        \centering\includegraphics[width=\textwidth]{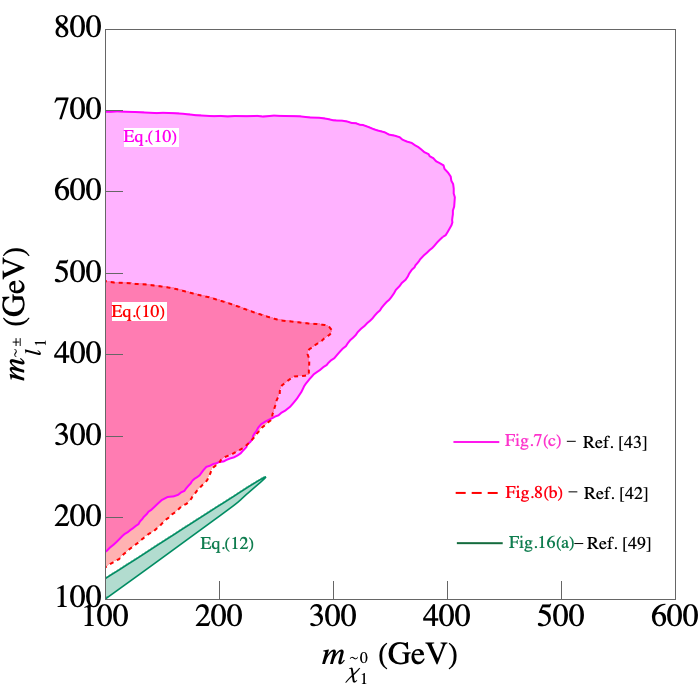}
                \caption{}
                \label{contour_mn1_mslep}
        \end{subfigure}
\caption{Latest LHC constraints in the 
$\mneu1$-$\mcha1$ (left) and $\mneu1$-$\msl1$ (right) planes (see text).
}
\label{lhccontours}        
\end{figure}

In this section we summarize the LHC Run-II results that are most relevant
for our analyses. A graphical ``comparison'' of the reach of these
various LHC constraints is shown in \reffi{lhccontours}, as described in
detail below.
In the left plot we show the different search modes in the
$\mneu1$-$\mcha1$ plane as given in the respective references.
In the right plot of \reffi{lhccontours}, we depict the limits on the
$\mneu1$-$\msl1$ plane \footnote{In the context of LHC searches, we
use leptons to imply only the first two generations unless otherwise
mentioned.}.
As will be argued in \refse{sec:scan} and
shown explicitly in \refse{sec:results} we focus on parameter ranges,
where either the light chargino or the sleptons are close in mass to $\neu1$
($\cha1$- or $\Slpm$-coannihilation).

\begin{itemize}
\item
{\bf{Decay via sleptons (\boldmath{$3l$}):}}~~

The pair production of electroweak gauginos decaying via
the process of \refeq{3ldecviaslep}
to $3 l$ final states is one of the most promising search
channels to look for electroweak SUSY particles at the LHC.
Exclusion limits from the ATLAS searches~\cite{Aaboud:2018jiw}
at 36~fb$^{-1}$ luminosity are based on a
``simplified model'' assuming that the decay of $\chapm1 / \neu2$
proceeds via an intermediate $\Sl_L$ 
or $\Sn_L$ with 100\% branching ratio, with $\mL = \msnu = (\mneu2 + \mneu1)/2$.
This search excludes upto $\mneu1 \sim 1100 \gev$ for $\mneu2 \lsim 550 \gev$.
However, it is sensitive to mass splittings $(\mcha1-\mneu1) \gtrsim 70 \gev$.
Thus, this limit is not sensitive to $\chapm1$-coannihilation region 
which corresponds to much smaller mass-splitting between $\chapm1$ and $\neu1$.
We impose this limit on the $\mcha1-\mneu1$ plane of
$\Sl$-coannihilation region. 
We show the exclusion contour from ATLAS in \reffi{contour_mn1_mch} as a
red dashed line.  
\begin{align}
\chapm1 \neu2 \to (\Sl^\pm \nu) (\Sl^{+} l^-) \to 3l + \met~, \notag \\
\chapm1 \neu2 \to (l^\pm \Sn) (\Sl^{+} l^-) \to 3l + \met~.
\label{3ldecviaslep}
\end{align}

\item
{\bf Decay via sleptons (\boldmath{$2l$}):}\\[.3em]
The decay chain in \refeq{chaviaslep} refers to the
ATLAS searches~\cite{Aad:2019vnb} at 139~fb$^{-1}$
for $\cha1$-pair production with subsequent decays through
$\Sl/\Sn$. The ATLAS analysis is performed for the specific parameter choice
$m_{\tilde l} = (\mneu2 + \mneu1)/2$ and is sensitive to
mass differences $\Delta m = (\mcha1 - \mneu1) \gtrsim 50 \gev$.
Thus, it is not possible to probe the $\cha1$-coannihilation region
with the help of these constraints.
As above, we impose these limits on the $\mcha1-\mneu1$ plane of the
$\Slpm$-coannihlation scenario.
In \reffi{contour_mn1_mch} this exclusion limit is shown as a magenta
line. 
\begin{align}
\chap1 \cham1 \to (\Sl^+ \nu) (\Sl^{-} \nu) \to 2l + \met~,
\notag \\
\chap1 \cham1 \to (l^+ \Sn) (l^{-} \Sn) \to 2l + \met~.
\label{chaviaslep}
\end{align}
The ATLAS limit for the same production and decay mode as in \refeq{chaviaslep}
at 36~fb$^{-1}$ luminosity given in \cite{Aaboud:2018jiw} is shown as a gray
dashed line in \reffi{contour_mn1_mch}. However, this provides a much weaker
limit for this scenario.
The CMS searches for gaugino-pair production with decay via sleptons/sneutrinos
to final states containing two or more leptons can be found
in \cite{Sirunyan:2017lae}.

\item
{\bf Decay via gauge bosons:}\\[.3em]
The ATLAS searches described by
\refeqs{3lviawz}-(\ref{chaviaww})~\cite{Aaboud:2018jiw,Aad:2019vnb}
look for decays of gaugino pairs through on-shell
gauge bosons. Thus, for these searches to be effective,
the mass difference between $\chapm1$ and $\neu1$ should at least be
$\sim$ electroweak scale, i.e.\ an on-shell gauge boson is required,
which makes them practically insensitive to the
$\chapm1$-coannihilation region.
On the other hand, these limits can in principle be applied
to the $\mcha1-\mneu1$ plane of the $\Slpm$-coannihlation region.
It should be noted that these searches are most effective for scenarios
where $\mL, \mR$ lie above $\mneu2, \mchap1$, so that gaugino-decay via
on-shell sleptons are not kinematically accessible. 

\citere{Aaboud:2018jiw} provides an exclusion contour at 36~fb$^{-1}$
combining the two channels given in \refeq{3lviawz} and \refeq{2ljetsviawz}.
This is by far the strongest limit for this kind of scenario.
We show the combined contour from ATLAS in \reffi{contour_mn1_mch}
by a cyan dashed line.
The various multi-lepton based searches are subdivided as:\\
a) $3l$:~~ The production and decay of the gauginos
are expected to proceed as in \refeq{3lviawz}.
\begin{subequations}\label{comb}
\begin{align}
\chapm1 \neu2 &\to (W \neu1) (Z \neu1) \to 3l + \met~,
\label{3lviawz}
\end{align}
b) $2l + jets$:~~ The search demands the presence in the signal
of two leptons of same flavour and opposite sign (SFOS) from
the $Z$-boson and two or more jets from hadronic decays
of the $W$, following \refeq{2ljetsviawz}.
\begin{align}
\chapm1 \neu2 &\to (W \neu1) (Z \neu1) \to 2l + jets + \met~.
\label{2ljetsviawz}
\end{align}
\end{subequations}

c) $2l$:~~ Apart from the above exclusion limit, there exists
ATLAS exclusion contours~\cite{Aad:2019vnb} at 139~fb$^{-1}$
for decay processes like \refeq{chaviaww} in final states
containing two opposite sign leptons. This is shown by
a blue line in \reffi{contour_mn1_mch}. However, this is much weaker
compared to the combined limit mentioned above.
\begin{align}
\chap1 \cham1 &\to (W^+ \neu1) (W^{-} \neu1) \to 2l + \met~.
\label{chaviaww}
\end{align}
Similar searches performed by the CMS collaboration are described in
\cite{Sirunyan:2018ubx}.

\item
{\bf Decay via Higgs boson (\boldmath{$l, b$}-jets):}\\[.3em]
There are searches performed by ATLAS~\cite{Aad:2019vvf}
 at 139~fb$^{-1}$ (and CMS~\cite{Sirunyan:2017zss} at 35.1~fb$^{-1}$) which 
look for decays of gaugino pairs through on-shell
gauge and Higgs bosons as in \refeq{chaviawh}.
These searches assume 100\,\% $\br$ for the decay mode $\neu2 \to h \neu1$, for a
purely wino-like $\neu2$. This limit is shown as a yellow-dashed line in \reffi{contour_mn1_mch}.
\begin{align}
\chapm1 \neu2 &\to (W \neu1) (h \neu1) \to l + b\bar b + \met~,
\label{chaviawh} 
\end{align}

\item
{\bf \boldmath{$\Sl$} pair production (\boldmath{$2l$}):}\\[.3em]
The ATLAS limit from the searches of slepton pair production in the
dilepton final state~\cite{Aad:2019vnb}  as described in \refeq{slepdec}
is sensitive for mass differences of
$(m_{\tilde l} - \mneu1) \gtrsim 80 \gev$,
making it inefficient for $\Slpm$-coannihilation region.
On the other hand, these limits can give constraints in
the $m_{\tilde l}-\mneu1$ plane of our $\cha1$-coannihilation
region. We show this as a magenta line in \reffi{contour_mn1_mslep}.
For the same search channel a weaker limit  at 36~fb$^{-1}$
luminosity comes from \citere{Aaboud:2018jiw}
which is shown as a red dashed line in \reffi{contour_mn1_mslep}.
Corresponding CMS search is described in \cite{Sirunyan:2018nwe}.

\begin{align}
\Sl^+ \Sl^{-} \to (l^+ \neu1) (l^- \neu1) \to 2l + \met~.
\label{slepdec}
\end{align}

\item
{\bf Compressed scenario:}\\[.3em]
Apart from the standard searches mentioned above, 
there are also dedicated analyses~\cite{Aad:2019qnd} to
investigate parameter spaces with compressed mass spectra, where the
mass-splittings between $\cha1,\neu2,\Slpm$ and $\neu1$ can be very low.
The ATLAS searches~\cite{Aad:2019qnd} at 139~fb$^{-1}$ are sensitive
to mass-splittings 
as low as $\sim$ 1.5 GeV for $\mneu2 \sim 100 \gev$ whereas for sleptons
the mass-splitting goes down to $\sim 550 \mev$ for $\msl{} \sim 70 \gev$.
In these cases, $\cha1$ and $\neu2$ are expected to
decay via off-shell $W$ and $Z$ bosons to final states containing
two leptons and $\met$ (i.e.\ requiring a leptonic $Z^*$~decay
and a hadronic $W^*$~decay), see \refeq{offshellwz},
whereas for the sleptons the decay
process is given in \refeq{compslepdec}.
In both cases, because of the small mass splitting, the two
same-flavor opposite-charge leptons in the final state come out to be
very soft. Thus, to increase the sensitivity of the searches,
the presence of initial-state radiation (ISR) is also required which gives
the system some amount of boost. The searches for
$\Sl^+\Sl^- (\cha1 \neu2)$-pair production can be applied to constrain
$\Sl~(\cha1)$-coannihilation regions.
These limits are shown in green both in \reffi{contour_mn1_mch}
and \reffi{contour_mn1_mslep}.
\cite{Sirunyan:2018iwl} describes the corresponding CMS searches
targeting the compressed mass spectrum.
\begin{align}
\chapm1 \neu2 \to (W^* \neu1) (Z^* \neu1)
              \to 2l + \met~+ {\rm ISR}~,
\label{offshellwz}\\
\Sl^+ \Sl^{-} \to (l^+ \neu1) (l^- \neu1)
             \to 2l + \met~+ {\rm ISR}~.
\label{compslepdec}
\end{align}

\item
{\bf Searches involving stau:}\\[.3em]
The ATLAS collaboration has looked for direct production of $\stau$-pairs
in final states involving hadronically decaying
$\tau$-leptons~\cite{Aad:2019byo}, 
yielding almost negligible limits for $\mneu1 > 100 \gev$. Although there
exists searches by the CMS collaboration~\cite{Sirunyan:2017lae} looking
for decays of $\chapm1$ and $\neu2$ to $\tau$-rich final states, there are
no similar searches by ATLAS so far.
Such limits are effective mostly
for parameter regions where decays of $\chapm1 / \neu2$ via first two
generations of sleptons are not kinematically allowed. 
As we will see in \refse{sec:paraana}, in our analysis 
the mass of the lighter stau lies closely below
$\msele1, \msmu1$ and/or $\msele2, \msmu2$
depending on the coannihilation scenario being considered.
Therefore the limits from the pair-production of the first two generations
of sleptons (\refeq{slepdec}) already put sufficiently
stringent limits on the parameter space relevant
for  stau searches.
Moreover, the CMS exclusion limits are sensitive only for mass differences
$(m_{\tilde \tau} - \mneu1) \gsim 30 \gev$.
Therefore, this limit will not be very sensitive
to parameter regions relying solely on stau-coannihilation mechanism
to achieve the right relic density value. 
Keeping these facts in mind we do not explicitly
take into account these limits in our analysis.

\end{itemize}

\begin{table}[t!]
\begin{center}
\begin{tabular}{l|ccccc}
\hline
Fig.              & Production \&   & Ref. & Luminosity   & Show in & \CM-\\
                  & decay mode      &      & (fb$^{-1}$)  & the color
                                                          & implementation\\
\hline
\multirow{6}{*}{\reffi{contour_mn1_mch}} & \refeq{3ldecviaslep} & \cite{Aaboud:2018jiw} & 36.1 & Red dashed & \cmark \\
                          & \refeq{comb} & \cite{Aaboud:2018jiw} & 36.1 & cyan dashed & \cmark\\
                          & \refeq{slepdec}     & \cite{Aaboud:2018jiw} & 36.1 & Gray dashed & \cmark \\             
                          & \refeq{chaviaslep} & \cite{Aad:2019vnb} & 139 & Magenta & \cmark\\
                          & \refeq{chaviaww}    & \cite{Aad:2019vnb} & 139 & Blue  & \cmark \\
                          & \refeq{compslepdec} & \cite{Aad:2019qnd} & 139 & Green & \xmark \\             
\hline
\multirow{3}{*}{\reffi{contour_mn1_mslep}} & \refeq{slepdec} & \cite{Aad:2019vnb} &  139 & Magenta & \cmark\\
                            & \refeq{slepdec} & \cite{Aaboud:2018jiw} & 36.1 & Red dashed & \cmark \\
                            & \refeq{offshellwz} & \cite{Aad:2019qnd} & 139 & Green  & \xmark \\
\hline
\end{tabular}
\caption{Table showing the relevant constraints from the LHC imposed to our
parameter space. 
The last column specifies which analyses have been
implemented by us into \CM.} 
\label{tablhc}
\end{center}
\end{table}

\noindent
The constraints imposed on the parameter space of our analysis
are summarized in \refta{tablhc} and represented ``graphically'' in 
\reffi{lhccontours} above.
As also discussed above,
the exclusion contours published by experimental collaborations
are based on ``simplified model'' approach.
However, for the SUSY scenarios discussed in our work, the kinematic configurations
as well as gaugino-compositions may be significantly different from
those analyses, leading to notable change in some of the exclusion limits.
Here, we use an independent implementation of the ATLAS analyses
\citere{Aaboud:2018jiw} and \citere{Aad:2019vnb}
with the use of the program package
\CM~\cite{Drees:2013wra,Kim:2015wza,Dercks:2016npn}. Below, we give a
brief description of our implementation. 

\smallskip
\CM\ is an analysis tool to test BSM scenarios against LHC constraints.
We implemented the searches of \citere{Aaboud:2018jiw} and \citere{Aad:2019vnb}
using the {\tt AnalysisManager} framework of \CM.
We validated our implementation against the cutflow
tables for various signal regions provided by the ATLAS collaboration.
For the validation, events with upto two additional partons were generated
using {\tt MadGraph5\_aMC@NLO}~\cite{Alwall:2014hca},
while showering and hadronization were performed with
{\tt PYTHIA8}~\cite{Sjostrand:2014zea}.
We used the PDF set {\tt NNPDF2.3LO}~\cite{Ball:2012cx} and CKKW-L
prescription~\cite{Lonnblad:2012ix} for jet-parton matching following the ATLAS analyses.
These events are then passed on to
\CM~which uses {\tt Fastjet}~\cite{Cacciari:2005hq, Cacciari:2011ma}
for jet reconstruction and {\tt Delphes}~\cite{deFavereau:2013fsa}
(which selects the ATLAS detector card) for fast detector simulation.
After the preparation of final detector-level objects, \CM~ performs a statistical
evaluation for every signal region of each analysis. It compares
the results of the simulation with the actual experimental data by computing
the parameter $r$ defined as,
\begin{equation}
r = \frac{S - 1.96 \times \Delta S}{S^{95}_{\rm exp}}
\end{equation}
where $S$ denotes the expected number of signal events, $\Delta S$ is
the $1~\sigma$ uncertainty on this number and $S^{95}_{\rm exp}$ is the 95\%
confidence level upper limit on the signal cross-section quoted by the
experimental collaboration. Thus, a parameter point is deemed to be excluded
by the analysis provided the value of $r$ exceeds 1.

Here, one should note that for the parameter region
where compressed searches~\cite{Aad:2019qnd} are most sensitive, the difference
between the ``simplified model'' and our scenario is rather mild.
Therefore we impose these constraints directly onto
our parameter space, keeping in mind that this provides a conservative
exclusion limit. 
On the other hand, as mentioned earlier, the
searches~\cite{Aad:2019vvf} described by \refeq{chaviawh} assume
100\,\% $\br$ for the decay mode $\neu2 \to h \neu1$ (implying
a predominantly wino-like $\neu2$).
In practice, however, the $\br$ for this decay mode is always
smaller than 100\% because of the presence of $\neu2 \to Z \neu1$ mode
\footnote{The $\br(\neu2 \to h \neu1)$ increases from $\sim$ 55\% at the
kinematic threshold $(\mneu2-\mneu1) \sim 125 \gev$ to $\sim 80\%$ for
slightly larger mass splittings for a predominantly wino-like $\neu2$.}
~(see also the discussion in \cite{Bharucha:2013epa,Han:2013kza}).
Thus, it is expected that this already sub-dominant limit will
effectively be even substantially weaker
for all practical SUSY scenarios. Therefore we do not apply this
limit and accordingly did not implement this
analysis independently in \CM.


\subsection{Constraints from measurements of \gmin2}
\label{sec:gmin2}

The experimental result for 
$\amu := \gmin2/2$ is dominated by the measurements made at the
Brookhaven National Laboratory (BNL)~\cite{Bennett:2006fi},
resulting in a world average of~\cite{PDG2018}
\begin{align}
\amu^{\rm exp} &= 11 659 209.1 (5.4) (3.3) \times 10^{-10}~,
\label{gmt-exp}
\end{align}
where the first uncertainty is statistical and the second systematic.
The SM prediction of \amu\ is given by~\cite{Keshavarzi:2019abf}%
\footnote{While completing this work a new ``world average'' SM value
appeared~\cite{Aoyama:2020ynm}, which is slightly lower
than \refeq{gmt-sm}, but also with a slightly larger uncertainty. Using
this value would have had a very small impact on our analysis.}
\begin{align}
\amu^{\rm SM} &= (11 659 181.08 \pm 3.78) \times 10^{-10}~.
\label{gmt-sm}
\end{align}
Comparing this with the current experimental measurement in \refeq{gmt-exp}
results in a deviation of~\cite{Keshavarzi:2019abf}
\begin{align}
\Delta\amu &= (28.02 \pm 7.37) \times 10^{-10}~, 
\label{gmt-diff}
\end{align}
corresponding to a $3.8\,\sig$ discrepancy.
This ``current'' result will be used below with a hard cut at 
$2\,\sig$ uncertainty.

Efforts to improve the experimental result at Fermilab by the
``MUON G-2'' collaboration~\cite{Grange:2015fou}
and at J-PARC~\cite{Mibe:2010zz} aim to reduce the experimental
uncertainty by a factor of four compared to the BNL measurement. 
For the second step in our analysis we consider the upcoming Run~1
result from the Fermilab experiment. 
The Run~1 data is expected to have roughly the same
experimental uncertainty as the current result in \refeq{gmt-exp}. 
We furthermore assume that the Run~1 data yields
the same central value as the current result. Consequently,
we anticipate that the combined uncertainty very
roughly shrinks by $1/\sqrt{2}$, yielding a future value of
\begin{align}
\De\amu^{\rm fut} & = (28.02 \pm 5.2) \times 10^{-10}~,
\label{gmt-fut}
\end{align}
corresponding to a $5.4\,\sig$ discrepancy.
Thus, the combination of Run~1 data with the existing
experimental \gmin2\ data has the potential to establish the
``discovery'' of BSM phyics.
This ``anticipated future'' result will be used below with a hard cut at 
$2\,\sig$ uncertainty.

\medskip
Recently a new lattice calculation for the leading order hadronic
vacuuum polarization (LO HVP) contribution to
$\amu^{\rm SM}$~\cite{Borsanyi:2020mff} has been reported.
Their claim is that using their
new lattice result, and in particular their strongly improved
uncertainty estimate, the discrepancy of the SM prediction with the
experimental result effectively disappears, in clear contrast to the
evaluations based on experimental
data~\cite{Keshavarzi:2019abf,Davier:2019can,Jegerlehner:2017lbd}. 
Subsequently, in \cite{Lehner:2020crt} it was argued that the
uncertainty estimate of the LO HVP contribution obtained
in \cite{Borsanyi:2020mff} is far too optimistic, confirming earlier
lattice based estimates.
Furthermore, in Ref.~\cite{Crivellin:2020zul} it was analyzed that the
lattice evaluation of the LO HVP contribution creates a severe tension
in the overall SM EW fit (see also \cite{Keshavarzi:2020bfy,1802677}). 
Consequently, while being aware of the theoretical developments of the
LO HVP contributions, we stick to the central value of the deviation of
$\amu$ as given in \refeq{gmt-diff} (in agreement
with~\cite{Aoyama:2020ynm}).
On the other hand, we are also
aware that our conclusions would change substantially if the result
presented in \cite{Borsanyi:2020mff} turned out to be correct.

\medskip
In MSSM the main contribution to \gmin2\ at the one-loop level comes from
diagrams involving $\cha1-\Sn$ and $\neu1-\tilde \mu$ loops. The
contributions are approximated
as~\cite{Moroi:1995yh,Martin:2001st,Badziak:2019gaf} 
\label{subsec:mssmgmin2}
\begin{eqnarray}
\label{amuchar}
a^{\tilde \chi^{\pm}-\Sn_{\mu}}_{\mu} &\approx&
\frac{\alpha \, m^2_\mu \, \mu\,M_{2} \tb}
{4\pi \sin^2\theta_W \, m_{\tilde{\nu}_{\mu}}^{2}}
\left( \frac{f_{\chi^{\pm}}(M_{2}^2/m_{\tilde{\nu}_{\mu}}^2)
-f_{\chi^{\pm}}(\mu^2/m_{\tilde{\nu}_{\mu}}^2)}{M_2^2-\mu^2} \right) \, ,
\\
\label{amuslep}
a^{\tilde \chi^0 -\tilde \mu}_{\mu} &\approx&
\frac{\alpha \, m^2_\mu \, \,M_{1}(\mu \tb-A_\mu)}
{4\pi \cos^2\theta_W \, (m_{\tilde{\mu}_R}^2 - m_{\tilde{\mu}_L}^2)}
\left(\frac{f_{\chi^0}(M^2_1/m_{\tilde{\mu}_R}^2)}{m_{\tilde{\mu}_R}^2}
- \frac{f_{\chi^0}(M^2_1/m_{\tilde{\mu}_L}^2)}{m_{\tilde{\mu}_L}^2}\right)\,,
\end{eqnarray}
where the loop functions $f$ are as given in Ref.~\cite{Badziak:2019gaf}.
In our analysis MSSM contribution to \gmin2\
at two loop order is calculated using {\tt
GM2Calc}~\cite{Athron:2015rva}, implementing two-loop corrections
from \cite{vonWeitershausen:2010zr,Fargnoli:2013zia,Bach:2015doa}
(see also \cite{Heinemeyer:2003dq,Heinemeyer:2004yq}).


\subsection{Dark matter relic density constraints}
\label{sec:relic}
The mean density of CDM in the Universe is tightly constrained by Planck
measurements of the cosmic 
microwave background and other observations~\cite{Planck}:
\begin{align}
\Omega_{\rm CDM} h^2 \; = \; 0.120 \pm 0.001 \, .
\label{OmegaCDM}
\end{align}

In an R-parity conserving scenario of MSSM, the lightest supersymmetric
particle (LSP) 
becomes a candidate for DM. Depending on the hierarchy among the parameters
$M_1, M_2$ and $\mu$ the LSP can be a bino-, wino- or higgsino-like state or
an admixture of them. It is well-understood that
with the current bounds on slepton masses from the LHC,
a predominantly bino-like LSP is unable to achieve correct amount of
relic density 
in the absence of some kind of coannihilation
mechanism~\cite{Go1983,ElHaNaOlSr1984}. 
A higgsino (wino) -like LSP, on the other hand is underabundant
upto the mass of $\sim$ 1 $\tev$ (3 $\tev$)~\cite{Hisano:2006nn,Cirelli:2007xd,
Hryczuk:2010zi,Beneke:2016ync,Bagnaschi:2016xfg}. Thus, in the absence
of sfermion-coannihilation, a well-tempered
admixture of higgsino-bino or higgsino-wino states can give rise to
the whole relic density of the universe single-handedly~\cite{ArkaniHamed:2006mb}.
We assume here that the dominant source of the cold DM is the LSP, the
lightest neutralino, $\neu1$, so that $\Och \simeq \Omega_{\rm CDM} h^2$. 
For the calculation of the DM relic density we use
\MO~\cite{Belanger:2001fz,Belanger:2006is,Belanger:2007zz,Belanger:2013oya}.


\subsection{Direct detection constraints of Dark matter}
\label{sec:dd}

We evaluate the constraint on the spin-independent
DM scattering cross-section $\ssi$ from
XENON1T~\cite{XENON} experiment, 
evaluating the theoretical prediction for $\ssi$ using
\MO~\cite{Belanger:2001fz,Belanger:2006is,Belanger:2007zz,Belanger:2013oya}.
While a combination  
with other DD experiments would yield slightly stronger limits, this
would have no relevant impact on our results, as will be discussed below.

The leading contribution to $\ssi$ in MSSM comes from t-channel Higgs exchange
and s-channel squark exchange diagrams. Since, in our analysis the squarks
and non-SM-like Higgses are taken to be significantly heavy, the contribution from
the lightest CP-even Higgs exchange will be the dominant one.
The $h-\neu1-\neu1$ coupling arises as a result of mixing
of the bino/wino and higgsino states~\cite{Hisano:2009xv}.
Thus, in the parameter region $\mu \simeq M_1, M_2$, the coupling can become
significantly large. Moreover, since the coupling is $\sim 1/\mneu1$,
a smaller $\ssi$ is expected for larger values of $\mneu1$.

The scenario of $\chapm1/\neu2$-coannihilation analyzed in our work
can arise from a mixed bino-wino
or a bino-higgsino type of LSP. A bino-dominated LSP, on the other hand
is expected to undergo $\Slpm$-coannihilation to achieve the right relic density.
Because of the tiny coupling of a bino with the Higgs, a predominantly-bino or
a bino-wino mixed DM tend to have rather small $\ssi$, often going below
the neutrino floor~\cite{floor,Duan:2018rls}.
However, as the higgsino component in the LSP increases, its
$\ssi$ can become large, thus receiving stringent bounds
from the direct detection experiments. However, we must note that
even a bino-higgsino well-tempered LSP can evade the direct detection bounds
in some fine-tuned regions of parameter space dubbed as the 'blind spots',
where the $\neu1$-$\neu1$-$h$ coupling at tree level becomes exactly equal to
zero~\cite{Cheung:2012qy,Han:2016qtc,Crivellin:2015bva}, and the signal can go
below the neutrino floor~\cite{floor}.
There are in principle, constraints on $\ssd$ as
well~\cite{Aprile:2019dbj, Amole:2019fdf}.
However, these limits are in general weaker than the
$\ssi$ ones.


\section{Parameter scan and analysis flow}
\label{sec:paraana}

\subsection{Parameter scan}
\label{sec:scan}

We scan the relevant MSSM parameter space to obtain lower and {\it upper}
limits on the relevant neutralino, chargino and slepton masses.
In order to achieve the correct DM relic density, see \refse{sec:relic},
by the lightest neutralino, $\neu1$, some mechanism such as a specific
co-annhihilation or pole annihilation has to be active in the early
universe. At the same time $\mneu1$ must not be too high, such that the 
EW sector can provide the contribution required to bring the theory
prediction of $\amu$ into agreement with the experimental measurement,
see \refse{sec:gmin2}. 
The combination of these two requirements yields the following
possibilities. (The cases present a certain choice of favored
possibilities, upon which one can expand, as will briefly discussed
in \refse{sec:conclusion}.)

\begin{description}
\item
{\bf (A) \boldmath{$\chapm1$}-coannihilation region}\\
In order to achieve $\cha1$-coannihilation $\mneu1 \sim \mcha1$ is
required. This can be achieved by
\begin{enumerate}
\item $M_1 \lsim M_2~(<< \mu)$ (bino-like LSP, or mixed bino-wino LSP)
\item $M_1 \lsim \mu~(<< M_2)$ (bino-like LSP, or mixed bino-higgsino LSP)
\item $M_2 < M_1, \mu$ (wino-like LSP)
\item $\mu < M_1, M_2$ (higgsino-like LSP)
\end{enumerate}
It is known~\cite{Hisano:2006nn,Cirelli:2007xd,Hryczuk:2010zi,Beneke:2016ync,
Bagnaschi:2016xfg} that a wino-like (higgsino-like) LSP fulfilling the
relic density constraint, \refeq{OmegaCDM}, results in
$\mneu1 \sim 2.9 (1.1) \tev$, which yields a SUSY spectrum too heavy to
fulfil the \gmin2\ constraint. On the other hand,
the possibility of mixed bino-higgsino LSP is strongly constrained
by the DD experiments, as discussed in \refse{sec:dd}.
Consequently, we are left with the bino
or mixed bino-wino like LSP. We choose the parameters according to, 
\begin{align}
  100 \gev \leq M_1 \leq 1 \tev \;,
  \quad M_1 \leq M_2 \leq 1.1 M_1\;, \notag \\
  \quad 1.1 M_1 \leq \mu \leq 10 M_1, \;
  \quad 5 \leq \tb \leq 60, \; \notag\\
  \quad 100 \gev \leq \mL \leq 1 \tev, \; 
  \quad \mR = \mL~.
\label{cha-coann}
\end{align}
Here we choose one soft SUSY-breaking parameter for all sleptons
together. While this choice should not have a relevant effect in the
$\cha1$-coannihilation case, this have an impact in the next case.
In our scans we will see that the chosen lower and upper limits are not
reached by the points that meet all the experimental constraints. This
ensures that the chosen intervals indeed cover all the relevant
parameter space. 

\item
{\boldmath{$\Slpm$}{\bf -coannihilation region}}\\
Another well-known mechanism to bring the relic density of the $\neu1$
into agreement with the experimental data is slepton coannihilation. As
above we choose only one soft SUSY-breaking parameter for 
all slepton generations.
This links automatically, stau-coannihilation and $\amu$, which in
principle are unrelated, see, e.g., \cite{deVries:2015hva,Bagnaschi:2017tru}. 
However, to keep the number of free parameters at a manageable level, we
keep this restriction in our analysis and leave the case with different
possible masses for different generations for future work.
On the other hand, we cover the two distinct cases that either the SU(2)
doublet sleptons, or the singlet sleptons are close in mass to the LSP.\\
{\bf (B)} Case-L: SU(2) doublet
\begin{align}
  100 \gev \leq M_1 \leq 1 \tev \;,
  \quad M_1 \leq M_2 \leq 10 M_1 \;, \notag\\
  \quad 1.1 M_1 \leq \mu \leq 10 M_1, \;
  \quad 5 \leq \tb \leq 60, \; \notag\\
  \quad M_1 \gev \leq \mL \leq 1.2 M_1, 
  \quad M_1 \leq \mR \leq 10 M_1~. 
\label{slep-coann-doublet}
\end{align}

{\bf (C)} Case-R: SU(2) singlet
\begin{align}
  100 \gev \leq M_1 \leq 1 \tev \;,
  \quad M_1 \leq M_2 \leq 10 M_1 \;, \notag \\
  \quad 1.1 M_1 \leq \mu \leq 10 M_1, \;
  \quad 5 \leq \tb \leq 60, \; \notag\\
  \quad M_1 \gev \leq \mR \leq 1.2 M_1,\; 
  \quad M_1 \leq \mL \leq 10 M_1~.
\label{slep-coann-singlet}
\end{align}
\end{description}
In all three scans we choose flat priors of the parameter space and
generate \order{10^7} points.

\begin{figure}
  \centering
  \begin{subfigure}[b]{0.48\linewidth}
    \centering\includegraphics[width=1.0\textwidth]{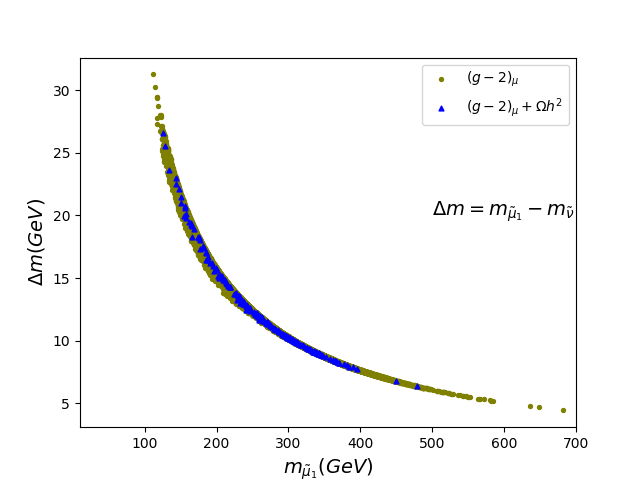}
    \caption{}
    \label{}
  \end{subfigure}
  ~
  \begin{subfigure}[b]{0.48\linewidth}
    \centering\includegraphics[width=1.0\textwidth]{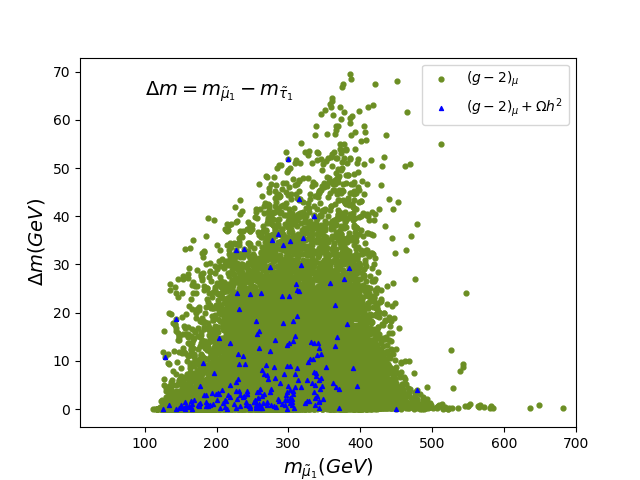}
    \caption{}
    \label{}
  \end{subfigure}
  \caption{The results of our parameter scan in the $\Delta-m_{\tilde \mu_1}$
    plane for the $\Slpm$-coannihilation
   Case-L. 
   .}
  \label{fig:Delta}
\end{figure}

In particular in the Case-L up to six sleptons can be close in mass, the
three charged ``left-handed'' sleptons as well as their respective
neutralinos. To give 
an idea of the still present mass splitting we show
in \reffi{fig:Delta} the mass difference between the light smuon
and (left) the muon sneutrino, or (right) the light stau.
In green we show the points fulfilling the \gmin2\ constraint (\refeq{gmt-diff}),
in dark blue the points that additionally give the correct DM relic density. 
The SU(2) relation enforces that the sneutrino is slightly lighter than
the light smuon, with a mass difference ranging from $\sim 25 \gev$ for
light smuons to about $\sim 5 \gev$ for heavy smuons. The non-zero
splitting in the scalar tau sector, for equal soft SUSY-breaking
parameters, makes the stau somewhat lighter than the smuon. Here the
mass difference can go up to $\sim 50 \gev$. Consequently, depending on
the parameter point, the NLSP can either be a sneutrino (the mass
differences between the three generations are negligible) or the light
scalar tau. 
For Case-R (not shown), taking also the DD limits into account, the
``left-handed'' sleptons being heavy, $\Stau2$ remains as 
the NLSP.

The mass parameters of the colored sector have been set to high values, such that
the resulting SUSY particle masses are outside the reach of the LHC, the
light $\cp$-even Higgs-boson is in agreement with the LHC measurements,
where the concrete values are not relevant for our analysis. $\MA$ has
also been set to be above the TeV scale. Consequently, we
do not include explicitly the possibility of $A$-pole annihilation,
with $\MA \sim 2 \mneu1$. As we will discuss below the combination of
direct heavy Higgs-boson searches with the other experimental
requirements constrain this possibility substantially (see,
however, also \refse{sec:conclusion}).
Similarly, we do not consider $h$-~or $Z$-pole annihilation, as such a
light neutralino sector likely overshoots the \gmin2\ contribution
(see, however, the discussion in \refses{sec:future-ee}
and \ref{sec:conclusion}). 

It should be kept in mind that while our three scenarios are
``designed'' to yield a certain coannihilation mechanism, the scan
over the EW parameters is quite extensive. This ensures in particular
that no ``possibilities'' to generate the required contribution
to \gmin2\ are overlooked.

\subsection{Analysis flow}
\label{sec:flow}

The data sample is generated by scanning randomly over the input parameter
range mentioned above, using a flat prior for all parameters.
We use {\tt SuSpect}~\cite{Djouadi:2002ze}
as spectrum and SLHA file generator. The points are required
to satisfy the $\chapm1$ mass limit from LEP~\cite{lepsusy}. 
The SLHA output files
from {\tt SuSpect} are then passed as input to {\tt GM2Calc} and \MO~for
the calculation of \gmin2 and the DM observables, respectively. The parameter
points that satisfy the current \gmin2\
constraint, \refeq{gmt-diff}, the DM
relic density, \refeq{OmegaCDM}, and the direct detection constraints are
then taken to the final 
step to be checked against the latest LHC constraints implemented in \CM,
as described in \refse{sec:collider}.%
\footnote{
We have also checked our points surviving all other constraints against
vacuum stability, in particular taking into account the scalar tau
sector~\cite{Endo:2013lva}, which can become relevant for large $\mu\tb$.
We found that all our surviving points pass this constraint.
}%
~The branching ratios of the relevant SUSY
particles are computed using {\tt SDECAY}~\cite{Muhlleitner:2003vg}
and given as input to \CM.



\section{Results}
\label{sec:results}


\subsection{\boldmath{$\cha1$}-coannihilation region}
\label{sec:chaco}

We start our discussion with $\cha1$-coannihilation, as discussed
in \refse{sec:scan}. We follow the analysis flow as described
in \refse{sec:flow} and denote the points surviving certain constraints
with different colors.  
\begin{itemize}
\item grey (round): all scan points.
\item green (round): all points that are in agreement with \gmin2, taking
into account the current or anticipated future limits,
see \refeqs{gmt-diff} and (\ref{gmt-fut}), respectively.
\item blue (triangle): points that additionally give the correct relic density,
see \refeq{OmegaCDM}.
\item cyan (diamond): points that additionally pass the DD constraints,
see \refse{sec:dd}.
\item red (star):  points that additionally pass the LHC constraints, see \refse{sec:collider}.
\end{itemize}

\begin{figure}[htb!]
\centering
\begin{subfigure}[b]{0.48\linewidth}
	\centering\includegraphics[width=\textwidth]{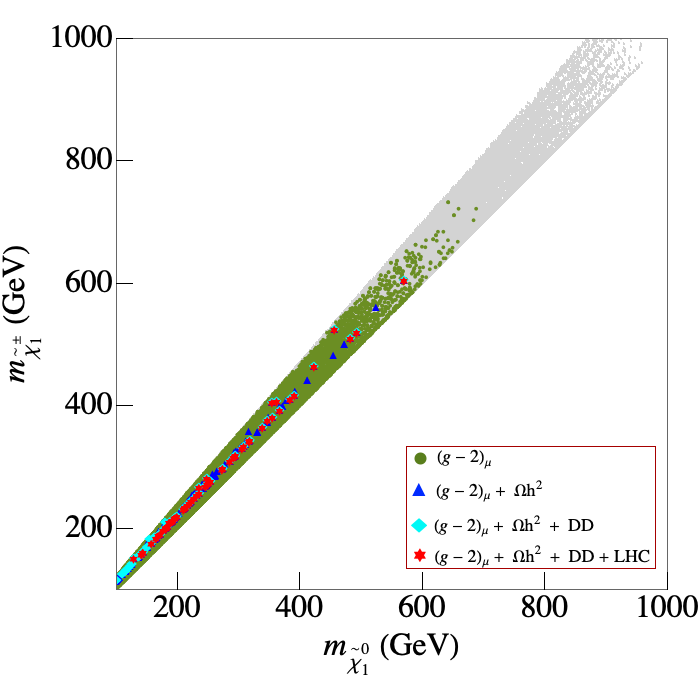}
	\caption{}
	\label{}
\end{subfigure}
~
\begin{subfigure}[b]{0.48\linewidth}
	\centering\includegraphics[width=\textwidth]{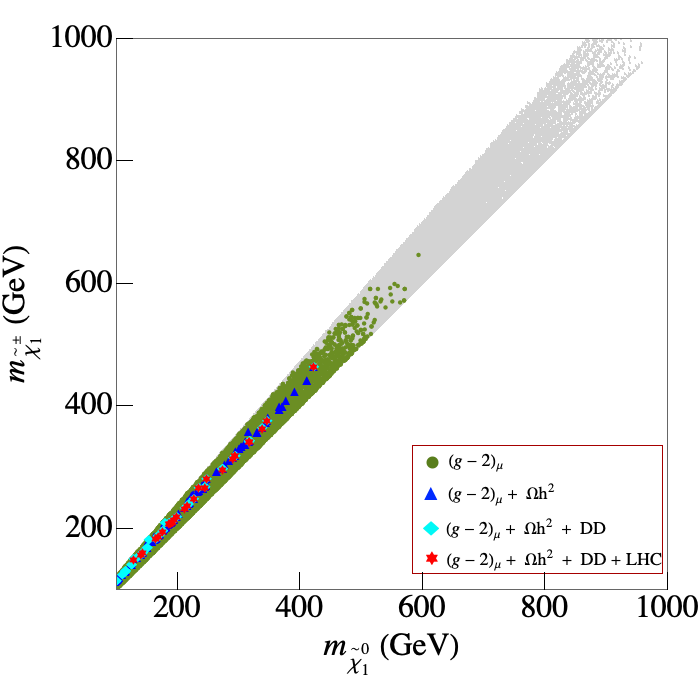}
	\caption{}
	\label{}
\end{subfigure}
\caption{The results of our parameter scan in the $\mneu1-\mcha1$ plane
for the $\cha1$-coannihilation scenario (current (left)
and anticipated future limits (right) from \gmin2). For the color coding: see text.
}
\label{mn1_mc1_chaco}
\end{figure}

In \reffi{mn1_mc1_chaco} we show our results in the $\mneu1$--$\mcha1$
plane for the current (left) and future (right) \gmin2\ constraint,
see \refeqs{gmt-diff} and (\ref{gmt-fut}), respectively. By definition
of $\cha1$-coannihilation the points are clustered in the diagonal of
the plane. Starting with the \gmin2\ constraint (green points) one can
observe a clear upper limits from \gmin2\ of about $700 \gev$ for the
current limits and about $600 \gev$ from the anticipated future
accuracy. Applying the CDM constraints reduce the upper limit further to
about $600 \gev$ and $450 \gev$, respectively. Applying the LHC
constraints, corresponding to the ``surviving'' red points (stars), does
not yield a further reduction from above, but cuts always (as
anticipated) only points in the lower mass range, see the discussion below.
The LHC constraint which is effective in this parameter plane is the
one designed for compressed spectra as given in \refeq{offshellwz} and
shown as a green line in \reffi{contour_mn1_mch}.
Thus, the experimental data set an upper as well as a lower bound,
yielding a clear search target for the upcoming LHC runs, and in
particular for future $e^+e^-$ colliders, as will be discussed
in \refse{sec:future}. In particular, this collider target gets
(potentially) sharpened substantially by the improvement in the \gmin2\
measurements.

\begin{figure}[htb!]
\vspace{2em}
\centering
\begin{subfigure}[b]{0.48\linewidth}
	\centering\includegraphics[width=\textwidth]{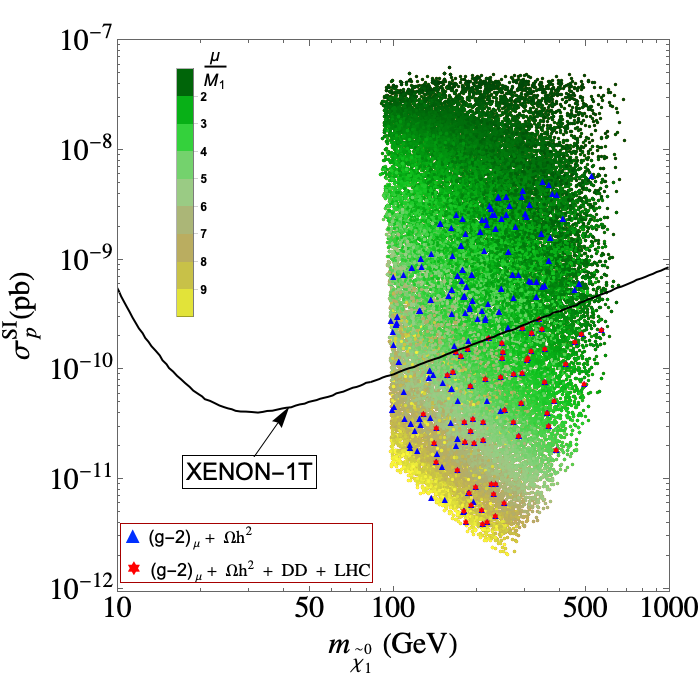}
	\caption{}
	\label{}
\end{subfigure}
~
\begin{subfigure}[b]{0.48\linewidth}
	\centering\includegraphics[width=\textwidth]{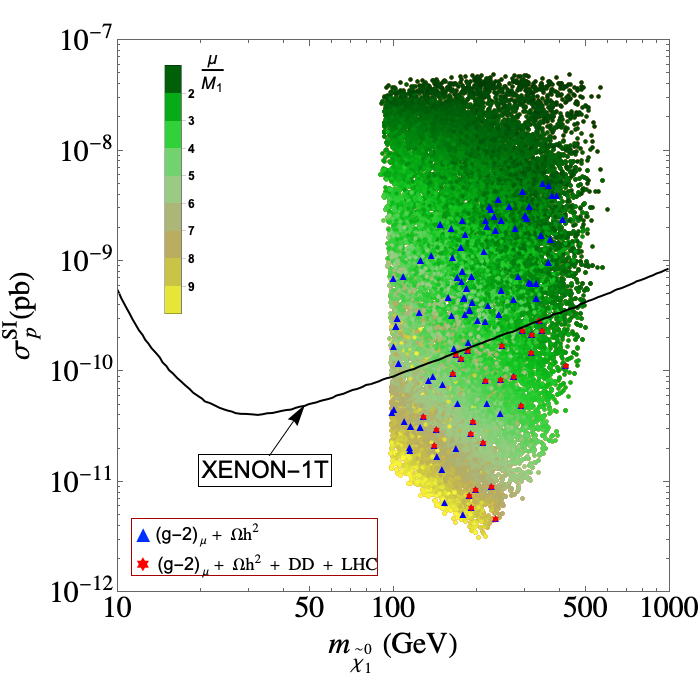}
	\caption{}
	\label{}
\end{subfigure}
\caption{Scan results in the $\mneu1$-$\ssi$ plane for current (left)
and anticipated future limits (right) from \gmin2. The color coding of
the points denotes $\mu/M_1$ and the black line indicates the DD
limits (see text). In blue (red) we show the points fulfilling the relic
density (and additionally the LHC) constraints.
}
\label{mneu1-ssi-chaco}
\end{figure}

The impact of the DD experiments is demonstrated 
in \reffi{mneu1-ssi-chaco}. We show the $\mneu1$-$\ssi$ plane for
current (left) and anticipated future limits (right) from \gmin2. 
The color coding of the points (from yellow to dark green) denotes $\mu/M_1$,
whereas in blue (red) we show the points fulfilling the relic
density (and additionally the LHC) constraints.
The black line indicates the current DD limits, here
taken for sake of simplicity from Xenon1T~\cite{XENON}, as discussed
in \refse{sec:dd}. It can be seen that a slight downward shift of
this limit, e.g.\ due to additional DD experimental limits from
LUX~\cite{LUX} or PANDAX~\cite{PANDAX}, would not
change our results in a relevant way.
The scanned parameter space extends from large $\ssi$
values, given for the smallest scanned $\mu/M_1$ values to the
smallest ones, reached for the largest scanned $\mu/M_1$, i.e.\
the $\ssi$ constraints is particularly strong for small $\mu/M_1$. 
One can also see that the relic density constraint is
fulfilled in nearly the whole scanned parameter space, except for the
largest $\ssi$ values. Given both CDM constraints and the LHC
constraints, shown in red, the smallest $\mu/M_1$ value we find
is~2 for the current and~2.3 for the
anticipated future \gmin2\ bound.
This result depends mildly on the
assumed \gmin2\ constraint, as this cuts away the largest $\mneu1$
values.

\begin{figure}[htb!]
\centering
\begin{subfigure}[b]{0.48\linewidth}
	\centering\includegraphics[width=\textwidth]{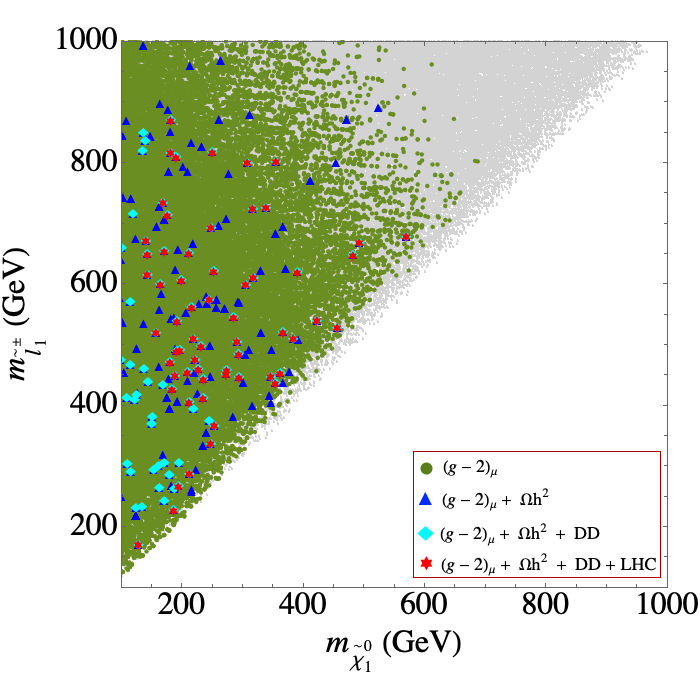}
	\caption{}
	\label{}
\end{subfigure}
~
\begin{subfigure}[b]{0.48\linewidth}
	\centering\includegraphics[width=\textwidth]{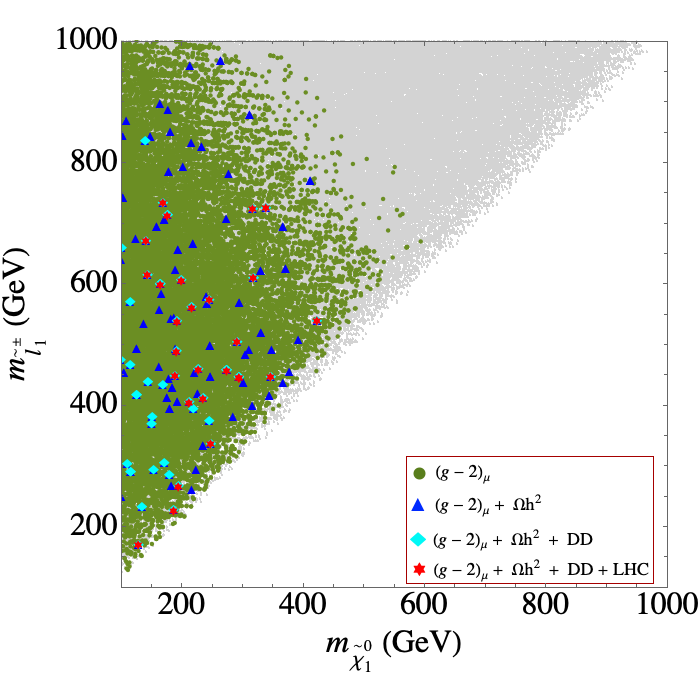}
	\caption{}
	\label{}
\end{subfigure}
\caption{The results of our parameter scan in the
$\mneu1$-$\msl1$ plane
in the $\chapm1$-coannihilation case (current (left)
and anticipated future limits (right) from \gmin2).
The color coding is as in \protect\reffi{mn1_mc1_chaco}.}
\label{mneu1-mslep-chaco}
\end{figure}

The distribution of the lighter slepton mass (where it should be kept in
mind that we have chosen the same masses for all three generations,
see \refse{sec:model}) is presented in the $\mneu1$-$\msl1$ plane
in \reffi{mneu1-mslep-chaco}, with the same color coding as
in \reffi{mn1_mc1_chaco}.
The \gmin2\ constraint places important constraints in this mass plane,
since both types of masses enter into the contributing SUSY diagrams,
see \refse{sec:gmin2}. 
The constraint is satisfied in a triangular region with its tip
around $(\mneu1, \msl1) \sim (700 \gev, 800 \gev)$ in the case of
current \gmin2\ constraints, and around $\sim (600 \gev, 700 \gev)$ in
the case of the anticipated future limits, i.e.\ the impact of the
anticipated improved limits is clearly visible as an {\it upper} limit. 
Since no specific other requirement is placed on the slepton sector in the
$\cha1$-coannihilation case the slepton masses are distributed over the
\gmin2\ allowed region. This also holds when the DM constraints are
taken in to account, as can be seen in the distribution of
the blue and cyan points (triangle/diamond).

The LHC constraints cut out lower slepton masses, going up to
$\msl1 \lsim 400 \gev$, as well as part of the very low $\mneu1$
points nearly independent of $\msl1$. Here the latter ``cut'' is due to the
searches for compressed spectra in \refeq{offshellwz}, shown as green line in
\reffi{contour_mn1_mch}. The first 
``cut'' is mostly a result of the searches described in \refeq{slepdec}.
It can be compared to the ``naive'' bound as given by the
magenta line in  \reffi{contour_mn1_mslep}, where this bound
extended up to $\msl1 \lsim 700 \gev$, a strong reduction is
observed.
The reason for this reduction can be understood as follows.
The ATLAS exclusion contour~\cite{Aad:2019vnb} for the mode in \refeq{slepdec}
is derived under the assumption that $\tilde e_L, \tilde \mu_L$
and $\tilde e_R, \tilde \mu_R$ are all mass-degenerate
and sufficiently light to contribute to the signal
cross-section. However, the right-sleptons can be significantly 
heavier for large parts of our parameter scan, with a
correspondingly reduced production cross-section as compared to the
ATLAS analysis. Moreover, $\tilde e_L$ has
significant $\br(\tilde e_L \ra \chapm1 \nu_e)$
and similarly for $\tilde \mu_L$. This also reduces the number
of signal leptons. Thus, a combination of these two factors leads to a
substantially weaker exclusion limit.
To illustrate this point better, in \refta{extab1} we show the mass spectra
and the relevant $\br$s of two representative points taken from the parameter
space of $\chapm1$-coannihilation scenario.

\begin{table}[!ht]
\small{
\begin{minipage}[t]{0.4\textwidth}
\begin{center}
\hspace{-1 cm}
\begin{tabular}[t]{|c||c|c|}
\hline
Sample point                   & 1  & 2    \\
\hline
\hline
$\mneu1$                       & 144   & 347  \\
\hline
$\mneu2$                       & 160   & 377   \\
\hline
$\mneu3$                       & 975  & 817    \\
\hline
$\mneu4 \sim \mcha2$           & 977 & 823   \\
\hline
$\mcha1$                       & 160  & 377 \\
\hline
$m_{\tilde e_1, \tilde \mu_1}$ & 617  & 449  \\
\hline
$m_{\tilde e_2, \tilde \mu_2}$ & 617   & 449  \\
\hline
$m_{\tilde \tau_1}$            &532 & 364 \\
\hline
$m_{\tilde \tau_2}$            &691 & 521 \\
\hline
$m_{\tilde \nu}$               &612 & 443 \\
\hline
\end{tabular}
 \end{center}
\end{minipage}
\begin{minipage}[t]{0.6\textwidth}
\begin{center}
\hspace{-1 cm}
\begin{tabular}[t]{|l|c|c||l|c|c|}
\hline
Sample point                      & 1   & 2   &     Sample point              & 1   &2 \\
\hline
\hline
$\br(\neu2 \ra \tilde \tau_1 \tau$ &  - &100  & $\br(\tilde e_1 \ra \neu1 e $ & 8.6    & 12.4 \\     
$\phantom{\br(\neu2 } \ra \neu1 \gamma$ & 12.7   & - &
$\phantom{\br(\tilde e_1 } \ra \neu2 e $   & 31.6    & 31.2 \\
$\phantom{\br(\neu2 } \ra \neu1 q \bar q$   & 22.5   & - &
$\phantom{\br(\tilde e_1 } \ra \chapm1 \nu_e) $ & 59.8   & 56.4 \\
$\phantom{\br(\neu2 } \ra \neu1 l \bar l$    & 17.6   & -    &                            &       &     \\
$\phantom{\br(\neu2 } \ra \neu1 \tau \bar\tau$ & 26.6   & -    &                            &       &     \\
$\phantom{\br(\neu2 } \ra \neu1 \nu \bar\nu$)  & 20.5   & -    &                            &       &     \\
\hline
$\br(\chapm1 \ra \neu1 \tau \nu_\tau$ & 10.8 & - &$\br(\tilde e_2 \ra \neu1 e $     & 99.9  & 99.5 \\
$\phantom{\br(\chapm1 } \ra \neu1 l \nu_l$       & 23     & - & $\phantom{\br(\tilde e_2 } \ra \neu2 e) $      & 0.1   & 0.5  \\
$\phantom{\br(\chapm1 } \ra \neu1 q \bar q'$    & 66      & - &                           &       &     \\
$\phantom{\br(\chapm1 } \ra \tilde \tau_1 \nu_\tau)$   & -      & 100 &                           &       &     \\
\hline
\end{tabular}
 \end{center}
\end{minipage}
\caption{The masses (in $\gev$) and relevant $\br$s (\%) of two representative points
from $\chapm1$-coannihilation
scenario. Here we show the \br\ of $\chapm1$ and $\neu2$ to third generation sleptons
separately and that of the first two generations together. Therefore,
$\Sl$ refers to $\tilde e$ and $\tilde \mu$ together. $\nu$ is used to
indicate $\nu_e$, $\nu_\mu$ and $\nu_\tau$ together.
}
\label{extab1}}
\end{table}
%
This emphasizes the importance of recasting using \CM, rather
than the ``naive'' application.
Overall we can place an upper limit on the light slepton
mass of about $\sim 850 \gev$ and $750 \gev$ for the current and the
anticipated future accurary of \gmin2, respectively. Since larger
values of slepton masses are reached for lower values of $\mneu1$, the
impact of \gmin2\ is relatively weaker than in the case of
chargino/neutralino masses. The phenomenological implications of
these limits will be discussed in \refse{sec:future}.

\begin{figure}[htb!]
	\vspace{3em}
\centering
\begin{subfigure}[b]{0.48\linewidth}
	\centering\includegraphics[width=\textwidth]{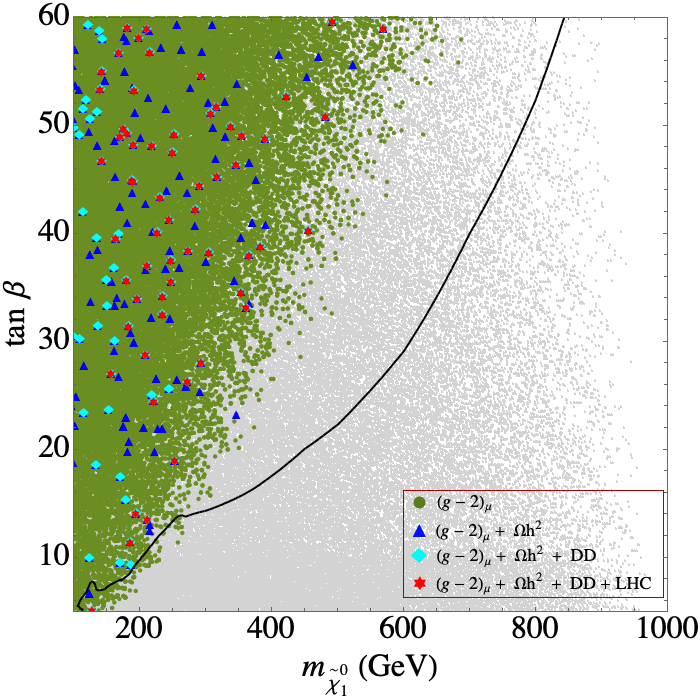}
	\caption{}
	\label{}
\end{subfigure}
~
\begin{subfigure}[b]{0.48\linewidth}
	\centering\includegraphics[width=\textwidth]{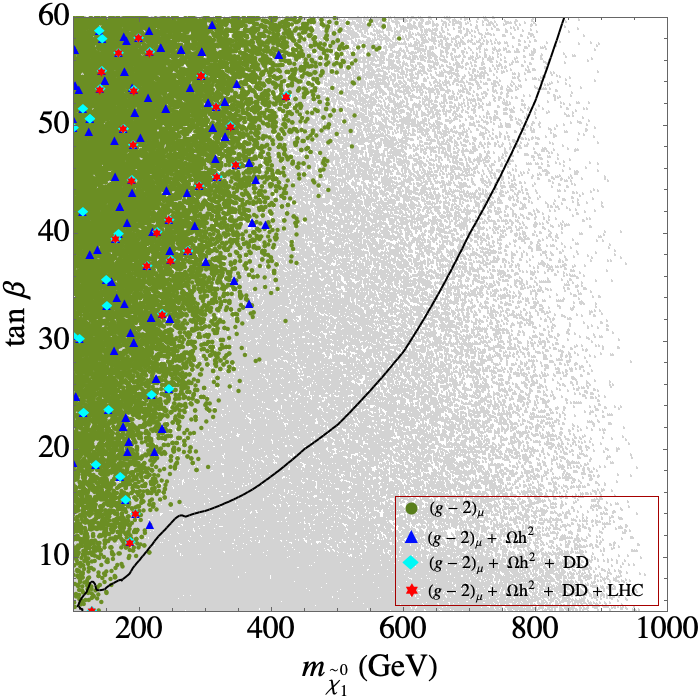}
	\caption{}
	\label{}
\end{subfigure}
\caption{The results of our parameter scan in the $\mneu1$-$\tb$ plane
for the $\chapm1$-coannihilation scenario (current (left)
and anticipated future limits (right) from \gmin2).
The color coding is as in \protect\reffi{mn1_mc1_chaco}.
The black line indicates the current exclusion bounds for heavy MSSM
Higgs bosons at the LHC (see text).}
\label{mneu1-tb-chaco}
\end{figure}

We finish our analysis of the $\cha1$-coannihilation case with the
$\mneu1$-$\tb$ plane presented in \reffi{mneu1-tb-chaco} with the same
color coding as in \reffi{mn1_mc1_chaco}. The \gmin2\ constraint is
fulfilled in a triangular region with largest neutralino masses allowed
for the largest $\tb$ values (where we stopped our scan at $\tb = 60$),
following the analytic dependence of the \gmin2\ contributions
in \refse{sec:gmin2}, $\amu \propto \tb/m_{\rm EW}^2$ (where we denote
with $m_{\rm EW}$ an overall EW mass scale.
In agreement with the previous plots, the largest values for the
lightest neutralino masses are $\sim 600 \gev$ $(\sim 450 \gev)$ for the
current (anticipated future) \gmin2\ constraint. 
The points allowed by the DM constraints (blue/cyan) are distributed
all over the allowed region. 
The LHC constraints cut out points at low $\mneu1$, but nearly
independent on $\tb$.

In \reffi{mneu1-tb-chaco} we also show as black lines the current bound
from LHC searches for heavy neutral Higgs
bosons~\cite{Bahl:2018zmf} in the channel $pp \to H/A \to \tau\tau$
in the $M_h^{125}(\tilde\chi)$ benchmark scenario
(based on the search data published
in \citere{Aaboud:2017sjh,Sirunyan:2018zut} using $36\, \ifb$.).%
\footnote{Stronger limits using $139\, \ifb$ have recently become
available~\cite{Aad:2020zxo}. However, no evaluation in the
$M_h^{125}(\tilde\chi)$ is available yet.}%
~In this scenario light
charginos and neutralinos are present, suppressing the $\tau\tau$ decay
mode and thus yielding relatively weak limits in the $\MA$-$\tb$ plane
(see Fig.~5 in~\cite{Bahl:2018zmf}). The black lines correspond to
$\mneu1 = \MA/2$, i.e.\ roughly to the requirement for $A$-pole
annihilation, where points
above the black lines are experimentally excluded. 
There are a few points passing the current \gmin2\ constraint
below the black $A$-pole line, reaching up to $\mneu1 \sim 280 \gev$,
for which the $A$-pole annihilation could provide the correct DM relic
density. It can be expected that with the improved limits as given
in \cite{Aad:2020zxo} this possibility is further restricted. Taking
into account the anticipated future \gmin2\ accuracy (keeping in mind the
hypothetical future central value is the same as the current one)
also cuts away most of the points below the black $A$-pole line. The
combination of these effects makes the $A$-pole annihilation in this
scenario marginal.


\subsection{\boldmath{$\Slpm$}-coannihilation region: Case-L}
\label{sec:case1}

We now turn to the case of $\Slpm$-coannihilation. As discussed in
in \refse{sec:scan} we distinguish two cases, depending which of the
two slepton soft SUSY-breaking parameters is set to be close to
$\mneu1$. In Case-L we chose $\msl{L} \sim M_1$, i.e.\ the left-handed
charged sleptons as well as the sneutrinos are close in mass to the LSP.
We find that all six sleptons are close in mass and differ by less
than $\sim 50 \gev$, see \reffi{fig:Delta}.

In \reffi{mn1_mse1_caseL} we show the results of our scan
in the $\msmu1 -\mneu1$ plane.
The color coding of the points is the same as in \reffi{mn1_mc1_chaco},
see the description in the beginning of \refse{sec:chaco}.
The green points in the left
plot satisfy the current \gmin2\ constraints of \refeq{gmt-diff}
whereas in the right plot these points correspond to the anticipated future
experimental bound of \gmin2\ as given in \refeq{gmt-fut}.
By definition of the scenario, the points are located along the
diagonal of the plane.
The present constraint from \gmin2\ puts an upper bound of
$\sim 650 \gev$ on the masses. With the projected sensitivity the upper
limit is slightly reduced to $\sim 550 \gev$. Including the DM and LHC
constraints, these bounds are reduced to $\sim 550 \gev$
and $\sim 500 \gev$ for the current and anticipated future \gmin2\
accuracy, respectively. 
As in the case of $\cha1$-coannihilation the LHC constraints cut away
only low mass points. 
The corresponding implications for the searches at future colliders
are discussed in \refse{sec:future}.

\begin{figure}[htb!]
\vspace{2em}
  \centering
  \begin{subfigure}[b]{0.48\linewidth}
    \centering\includegraphics[width=\textwidth]{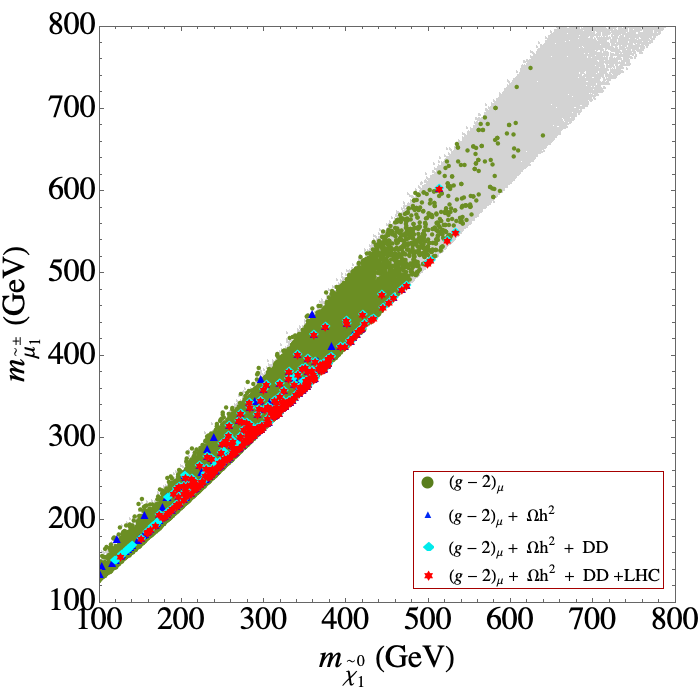}
    \caption{}
    \label{}
  \end{subfigure}
  ~
  \begin{subfigure}[b]{0.48\linewidth}
    \centering\includegraphics[width=\textwidth]{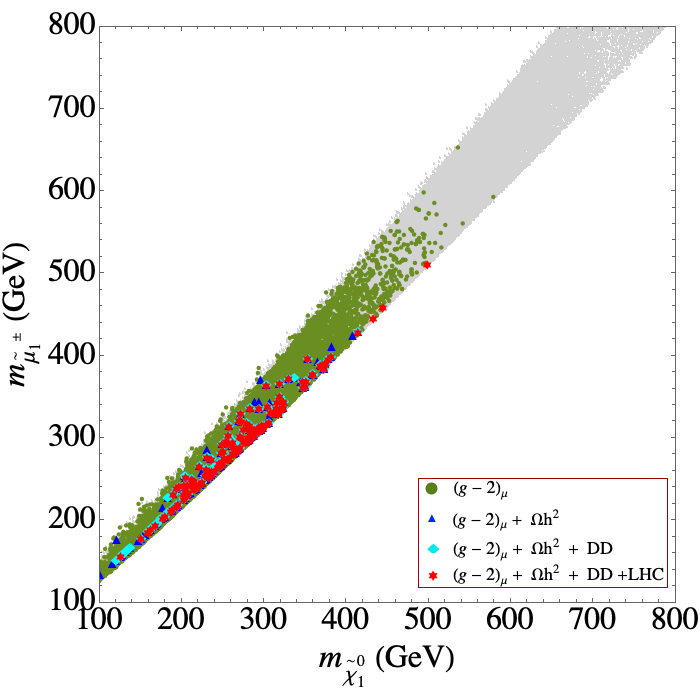}
    \caption{}
    \label{}
  \end{subfigure}
  \caption{The results of our parameter scan in the $\msmu1-\mneu1$ plane for the $\Slpm$-coannihilation
   Case-L (current (left)
and anticipated future limits (right) from \gmin2). The color coding as in \protect\reffi{mn1_mc1_chaco}.}
  \label{mn1_mse1_caseL}
\end{figure}

The impact of the DD experiments in the Case-L is demonstrated 
in \reffi{mneu1-ssi-caseL}. We show the $\mneu1$-$\ssi$ plane for
current (left) and anticipated future limits (right) from \gmin2. 
The color coding of the points is as in \reffi{mneu1-ssi-chaco}. 
As above, the black line indicates the current DD limits~\cite{XENON}.
The general features are as in the $\cha1$-coannihilation scenario:
the scanned parameter space extends from large $\ssi$
values, given for the smallest scanned $\mu/M_1$ values to the
smallest ones, reached for the largest scanned $\mu/M_1$.
One can also see that the relic density constraint is
fulfilled in nearly the whole scanned parameter space
but spreading mostly towards higher $\mu/M_1$ values.
Given both CDM constraints and the LHC
constraints, the smallest $\mu/M_1$ value we find is~2.46 for both the
current and the anticipated future \gmin2 bound.

\begin{figure}[htb!]
  \centering
  \begin{subfigure}[b]{0.48\linewidth}
    \centering\includegraphics[width=1.0\textwidth]{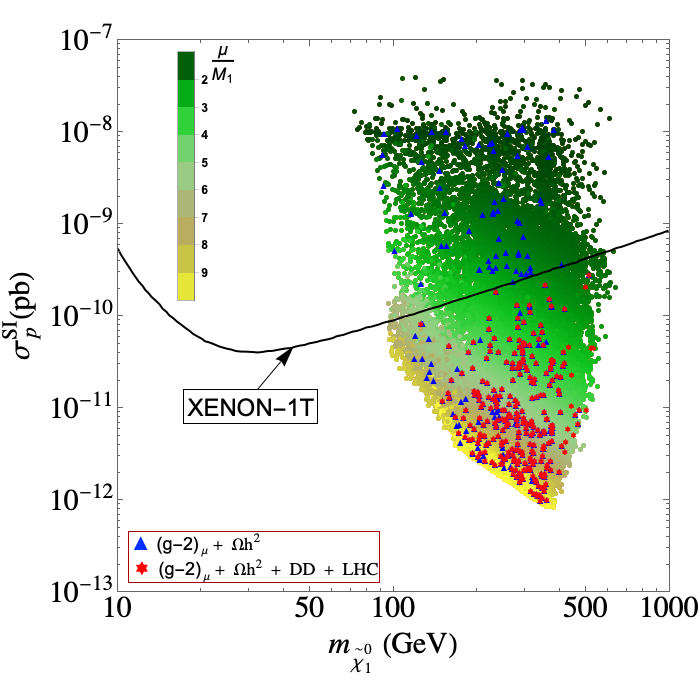}
    \caption{}
    \label{}
  \end{subfigure}
  ~
  \begin{subfigure}[b]{0.48\linewidth}
    \centering\includegraphics[width=1.0\textwidth]{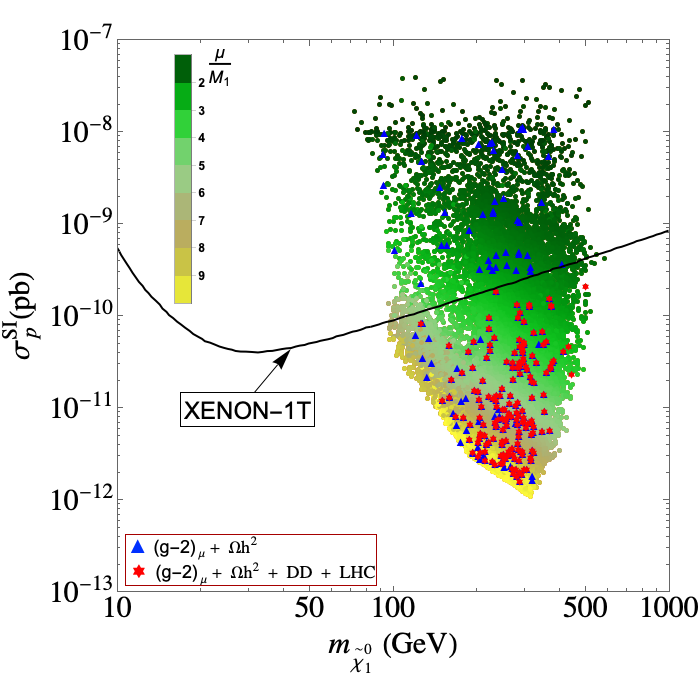}
    \caption{}
    \label{}
  \end{subfigure}
  \caption{The results of our parameter scan in the $\ssi-\mneu1$ plane
  for the $\Slpm$-coannihilation 
   Case-L (current (left)
and anticipated future limits (right) from \gmin2).
The color coding as in \protect\reffi{mneu1-ssi-chaco}.
}
  \label{mneu1-ssi-caseL}
\end{figure}

In \reffi{mneu1-mcha1-caseL} we show the results in the
$\mneu1$-$\mcha1$ plane with the same color coding as
in \reffi{mn1_mse1_caseL}. The \gmin2\ limits on $\mneu1$ become
slightly stronger for larger chargino masses, as expected
from \refeq{amuslep}, and upper limits on the chargino mass are set at
$\sim 3 \tev$ ($\sim 2.5 \tev$) for the current (anticipated future)
precision in $\amu$.
The LHC limits cut away a lower wedge going up to
$\mcha1 \lsim 600 \gev$, driven by the bound in \refeq{3ldecviaslep},
shown as the red dashed line in \reffi{contour_mn1_mch}.
As in the $\cha1$-coannihilation case, also here the upper limit on
$\mcha1$ is strongly reduced w.r.t.\ the ``naive'' application, which
goes up to $\mcha1 \lsim 1100 \gev$ for negligible $\mneu1$.
The reason for the weaker limit can be attributed to
two factors. First, the significant branching ratios of
$\br(\chapm1 \to \Stau1 \nu_\tau)$
and $\br(\neu2 \to \Stau1 \tau)$ respectively,
which are considered to be absent in the ATLAS analysis.
Second, the notably large branching ratio of
$\neu2$ to the invisible modes $\neu2 \to \Sn \nu$.
\refta{extab2} gives an idea of the relevant $\br$s of
two sample points taken from the parameter space of Case-L,
with their mass spectra given in the same table.
This again emphasizes the importance of the recasting of the LHC searches
that we have applied.

\begin{figure}
\vspace{2em}
  \centering
  \begin{subfigure}[b]{0.48\linewidth}
    \centering\includegraphics[width=\textwidth]{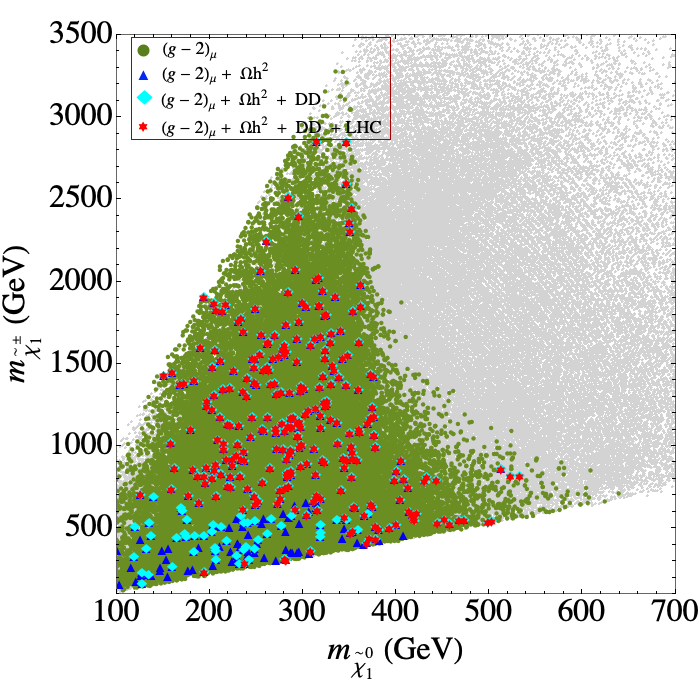}
    \caption{}
    \label{}
  \end{subfigure}
  ~
  \begin{subfigure}[b]{0.48\linewidth}
    \centering\includegraphics[width=\textwidth]{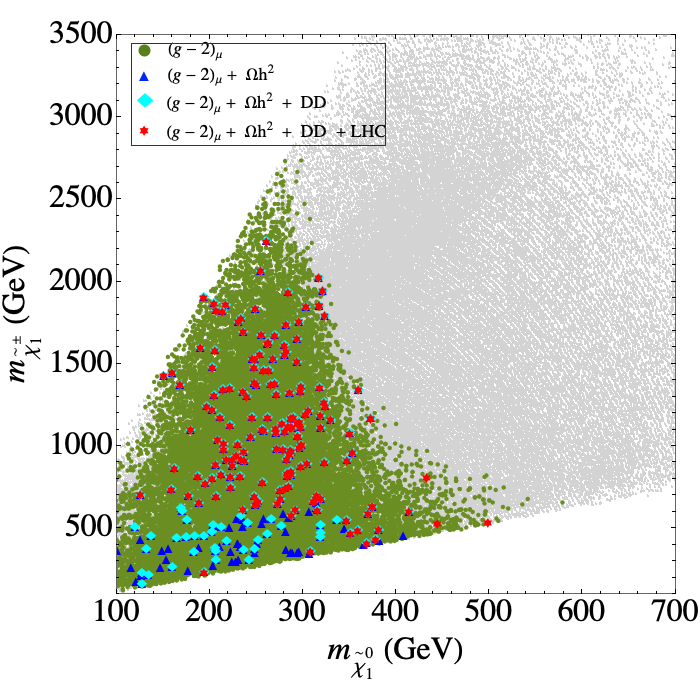}
    \caption{}
    \label{}
  \end{subfigure}
  \caption{The results of our parameter scan in the $\mcha1-\mneu1$
  plane for the $\Slpm$-coannihilation 
   Case-L (current (left)
and anticipated future limits (right) from \gmin2).
The color coding as in \protect\reffi{mn1_mse1_caseL}.}
\label{mneu1-mcha1-caseL}
\end{figure}

\begin{figure}[htb!]
\vspace{2em}
  \centering
  \begin{subfigure}[b]{0.48\linewidth}
    \centering\includegraphics[width=\textwidth]{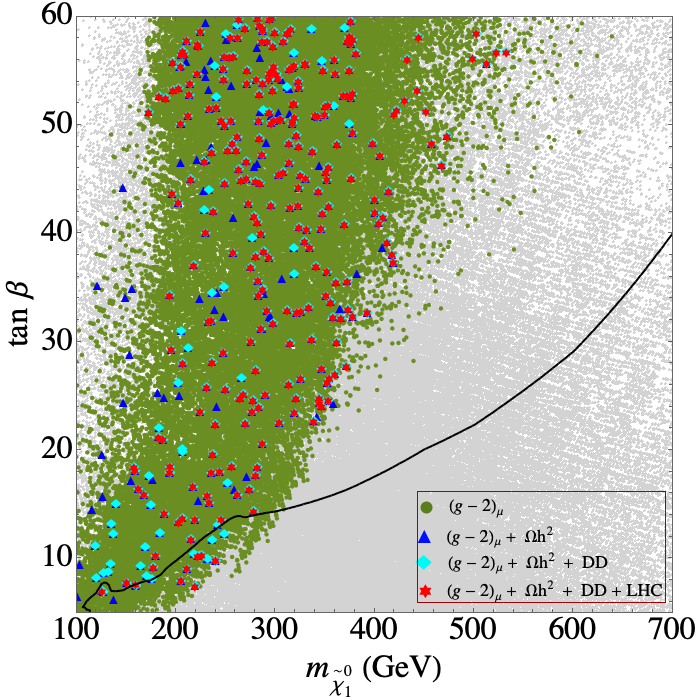}
    \caption{}
    \label{}
  \end{subfigure}
  ~
  \begin{subfigure}[b]{0.48\linewidth}
    \centering\includegraphics[width=\textwidth]{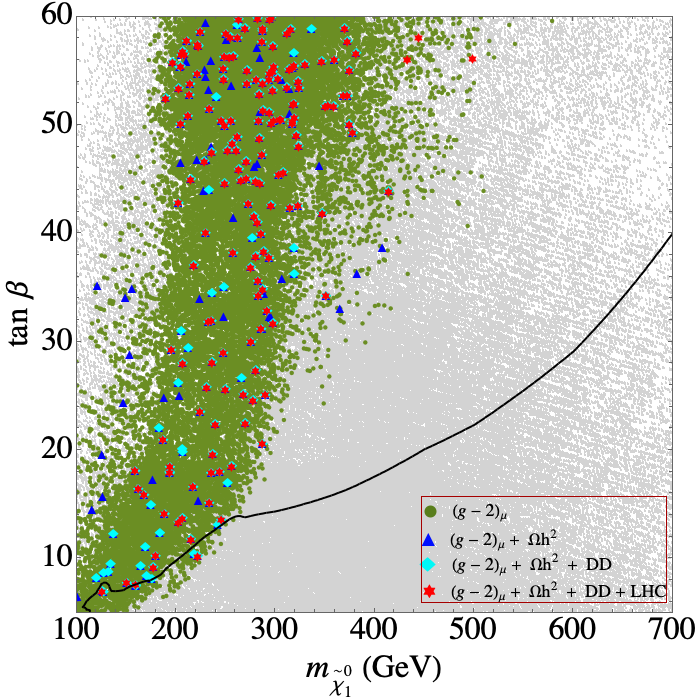}
    \caption{}
    \label{}
  \end{subfigure}
  \caption{The results of our parameter scan in the $\tb-\mneu1$ plane for the $\Slpm$-coannihilation
   Case-L (current (left)
and anticipated future limits (right) from \gmin2).
The color coding is as in \protect\reffi{mn1_mse1_caseL}.
}
\label{mneu1-tb-caseL}
\end{figure}

\begin{table}[!htb]
\begin{minipage}[t]{0.4\textwidth}
{\small
\begin{center}
\hspace{-1 cm}
\begin{tabular}[t]{|c||c|c|}
\hline
Sample point                   & 1    & 2   \\
\hline
$\mneu1$                       & 223 & 474 \\ 
\hline
$\mneu2$                       & 749 & 549      \\
\hline
$\mneu3$                       & 752  & 1934 \\
\hline
$\mneu4 \sim \mcha2$           & 1798  &1935 \\
\hline
$\mcha1$                       & 748   & 549 \\
\hline
$m_{\tilde e_1, \tilde \mu_1}$ & 243 & 484   \\
\hline
$m_{\tilde e_2, \tilde \mu_2}$ & 2137 & 4007 \\
\hline
$m_{\tilde \tau_1}$            & 240  & 482 \\
\hline
$m_{\tilde \tau_2}$            & 2137  & 4008 \\
\hline
$m_{\tilde \nu}$               & 231 & 478  \\
\hline
\end{tabular}
 \end{center}
}
\end{minipage}
\begin{minipage}[t]{0.6\textwidth}
{\small
\begin{center}
\hspace{-1 cm}
\begin{tabular}[t]{|l|c|c||l|c|c|}
\hline
Sample point                   & 1    & 2   & Sample point                   & 1    & 2   \\                                 
\hline
$\br(\neu2 \ra \Sl_1 l$         & 1.2   & 30.4 & $\br(\chapm1 \ra \tilde \nu_l l$     &  0.8 & 35.8\\ 
$\phantom{\br(\neu2} \ra \tilde \tau_1 \tau$  & 83.4  & 16.2 &$\phantom{\br(\chapm1} \ra \tilde \nu_{\tau_1} \tau$      &  84  & 18\\
$\phantom{\br(\neu2} \ra \tilde \nu \nu$ & -     & 53.3 &$\phantom{\br(\chapm1} \ra \Sl_1 \nu_l$               & 0.16 & 30 \\
$\phantom{\br(\neu2} \ra \neu1 h$             & 12    &  -   &$\phantom{\br(\chapm1} \ra \tilde \tau_1 \nu_\tau$    &  0.2 & 16\\ 
$\phantom{\br(\neu2} \ra \neu1 Z)$            & 3.4   &  -   &$\phantom{\br(\chapm1} \ra W \neu1)$                   & 14   & - \\ 
                                &       &      &                        &       &      \\
\hline
$\br(\tilde e_1 \ra \neu1 e) $  & 100   & 100  &$\br(\tilde e_2           \ra \neu1 e $ & 99.7 & 100\\
                                &       &      &$\phantom{\br(\tilde e_2} \ra \neu2 e $) & 0.3 & \\
                                &       &      &                        &       &      \\
                                &       &      &                        &       &      \\                                
\hline
\end{tabular}
 \end{center}
}
\end{minipage}
\caption{The masses (in $\gev$) and relevant $\br$s (\%) of two sample points
from $\Slpm$-coannihilation Case-L
scenario. Here we show the $\br$s of $\chapm1$ and $\neu2$ to third generation sleptons
separately and that of the first two generations together. Therefore,
$\Sl$ refers to $\tilde e$ and $\tilde \mu$ together. $\nu$ is used to
indicate $\nu_e$, $\nu_\mu$ and $\nu_\tau$ together.
}
\label{extab2}
\end{table}

The results for the $\Slpm$-coannihilation Case-L in the
$\mneu1$-$\tb$ plane are presented in \reffi{mneu1-tb-caseL}. The
overall picture is similar to the $\cha1$-coannhiliation case shown
above in \reffi{mneu1-tb-chaco}. Larger LSP masses are allowed for
larger $\tb$ values. On the other hand the combination of small $\mneu1$
and large $\tb$ leads to a {\it too large} contribution to
$\amu^{\rm SUSY}$ and is thus excluded. As in \reffi{mneu1-tb-chaco}
we also show the limits from $H/A$ searches at the LHC, where we set
(as above) $\mneu1 = \MA/2$, i.e.\ roughly to the requirement for $A$-pole
annihilation, where points above the black lines are experimentally excluded. 
In this case for the current \gmin2\ limit substantially more
points passing  the \gmin2\ constraint ``survive'' below the black
line, i.e.\ they are potential candidates for $A$-pole
annihilation. The masses reach up to $\sim 320 \gev$. As in the case
of $\cha1$-coannihilation, see \reffi{mneu1-tb-chaco}, these points
are reduced in the case of the anticipated future \gmin2\ accuracy
with an upper limit of $\sim 260 \gev$. Together with the already
stronger bounds on $H/A \to \tau\tau$~\cite{Aad:2020zxo} this does not
fully exclude $A$-pole annihilation, but leaves it as a rather remote
possibility with a clear upper bound on $\mneu1$ (see the discussion
in \refse{sec:future-ee}).



\subsection{\boldmath{$\Slpm$}-coannihilation region: Case-R}
\label{sec:case2}

We now turn to our third scenario, $\Slpm$-coannihilation Case-R,
where in the scan we require the ``right-handed'' sleptons to be close
in mass with the LSP. It should be kept in mind that in our notation
we do not mass-order the sleptons: for negligible mixing as it is
given for selectrons and smuons the ``left-handed'' (``right-handed'')
slepton corresponds to $\Sl_1$ ($\Sl_2$).
As it will be seen below, in this scenario all
relevant mass scales are required to be relatively light by the
\gmin2\ constraint. 

We start in \reffi{mneu1-msmu2-caseR} with the $\mneu1$-$\msmu2$ plane
with the same color coding as in \reffi{mn1_mc1_chaco}. By definition
of the scenario the points are concentrated on the diagonal. The
current (future) \gmin2\ bound yields upper limits on the LSP of
$\sim 700 (600) \gev$, as well as an upper limit on $\msmu2$ (which is close
in mass to the $\Sel2$ and $\Stau2$) of $\sim 800 (700) \gev$.
Including the CDM and LHC constraints, these limits reduce to
$\sim 520~(400) \gev$ for the LSP for the current (future) \gmin2\
bounds, and correspondingly to $\sim 600~(430) \gev$ for $\msmu2$,
and $\sim 530~(410) \gev$ for $\mstau2$.
The LHC constraints cut out some, but not all lower-mass
points. 

\begin{figure}[htb!]
	\vspace{2em}
	\centering
	\begin{subfigure}[b]{0.48\linewidth}
		\centering\includegraphics[width=\textwidth]{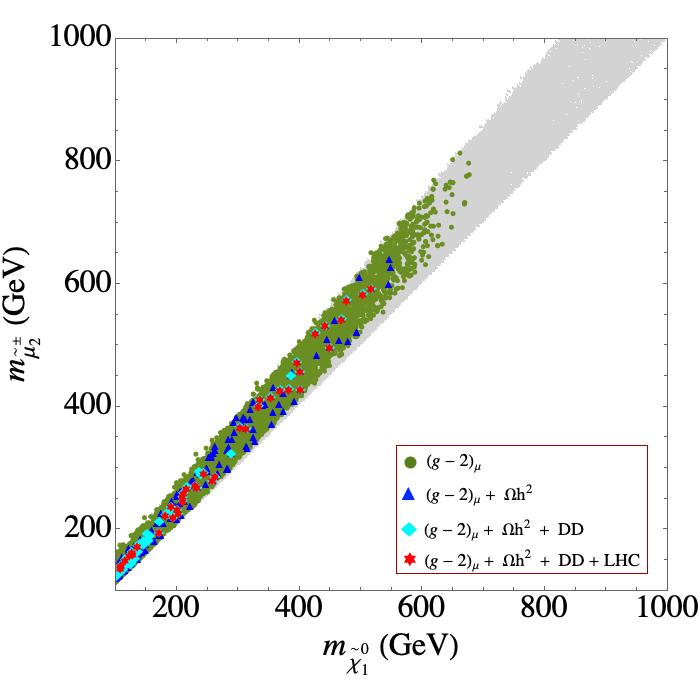}
		\caption{}
		\label{}
	\end{subfigure}
	~
	\begin{subfigure}[b]{0.48\linewidth}
		\centering\includegraphics[width=\textwidth]{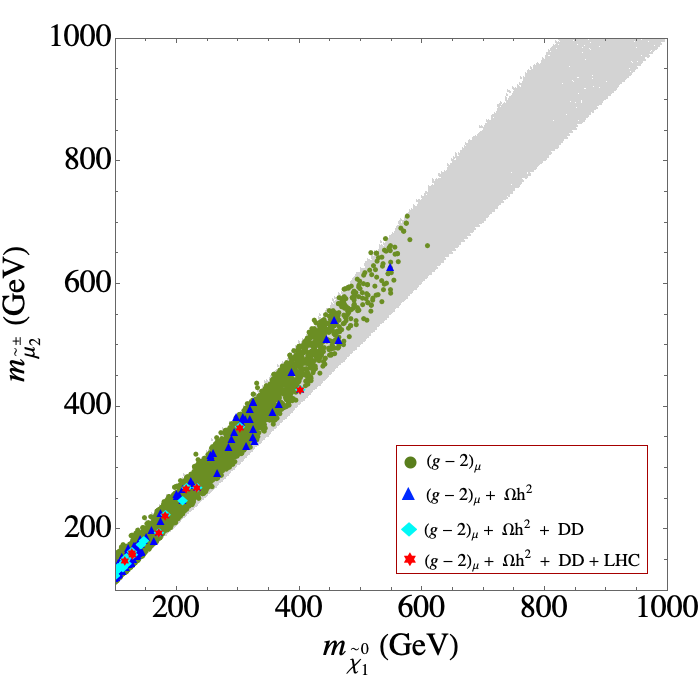}
		\caption{}
		\label{}
	\end{subfigure}
 \caption{The results of our parameter scan in the $\mneu1$-$\msmu2$
	plane for the $\Slpm$-coannihilation 
	Case-R (current (left)
and anticipated future limits (right) from \gmin2).
The color coding as in \protect\reffi{mn1_mc1_chaco}.
}
\label{mneu1-msmu2-caseR}
\end{figure}

\begin{figure}[htb!]
	\vspace{2em}
	\centering
	\begin{subfigure}[b]{0.48\linewidth}
		\centering\includegraphics[width=\textwidth]{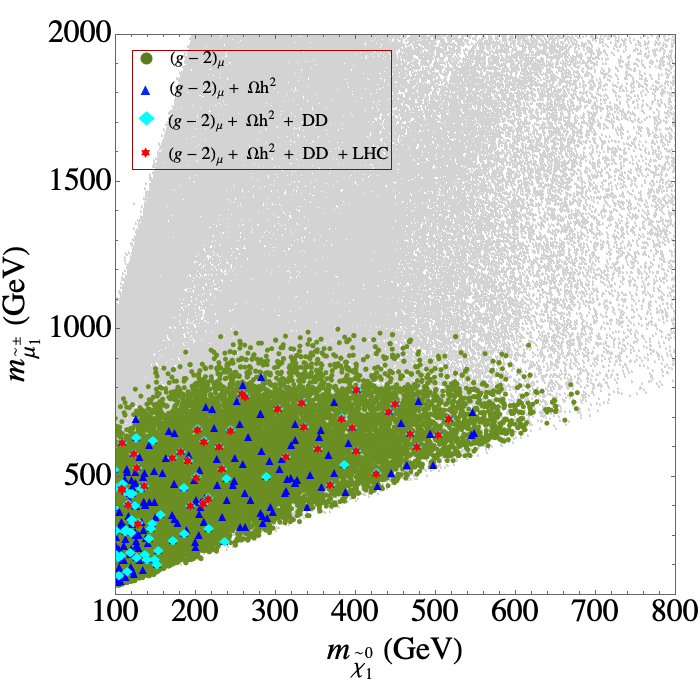}
		\caption{}
		\label{}
	\end{subfigure}
	~
	\begin{subfigure}[b]{0.48\linewidth}
		\centering\includegraphics[width=\textwidth]{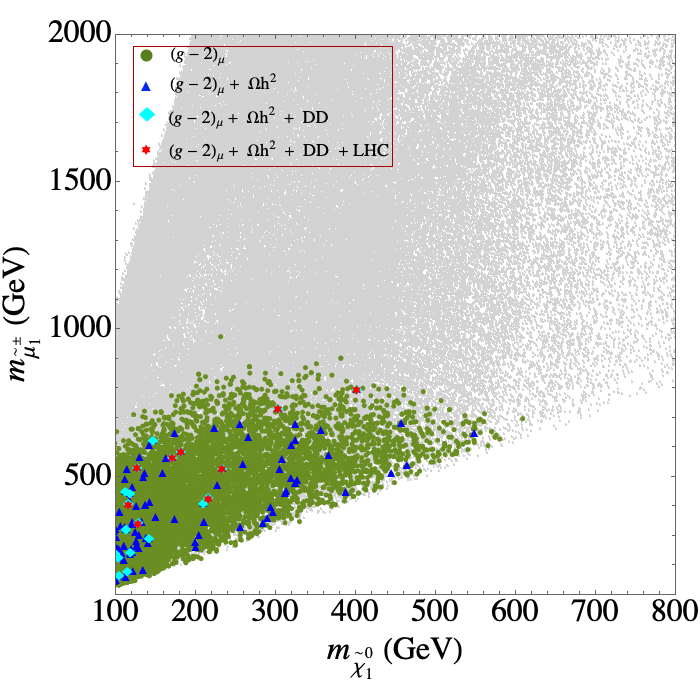}
		\caption{}
		\label{}
	\end{subfigure}
\caption{The results of our parameter scan in the $\mneu1$-$\msmu1$
	plane for the $\Slpm$-coannihilation 
	Case-R (current (left)
and anticipated future limits (right) from \gmin2).
The color coding as in \protect\reffi{mneu1-msmu2-caseR}.}
\label{mneu1-msmu1-caseR}
\end{figure}

The distribution of the heavier slepton is displayed in the
$\mneu1$-$\msmu1$ plane in \reffi{mneu1-msmu1-caseR}. Although the
``left-handed'' sleptons are allowed to be much heavier, the \gmin2\
constraint imposes an upper limit of $\sim 950~(800) \gev$ in the case
for the current (future) \gmin2\ precision. This can be understood from the SUSY contributions to $\amu$. If the ``left-handed''
slepton is heavy this suppresses both contributions, the
chargino-sneutrino loop as given in \refeq{amuchar}, as well as the
neutralino-slepton loop as given in \refeq{amuslep}. Consequently, in
particular the ``left-handed'' slepton soft SUSY-breaking parameter
must not be too large to give a relevant contribution to $\amu^{\rm SUSY}$. 
Valid parameter point with maximum $\msmu1$ $\sim 750 \gev$ for both the
current and future \gmin2\ precision are obtained,  even after the
inclusion of DM and LHC constraints.
This offers in principle very optimistic future collider
prospects, as will be discussed in \refse{sec:future}.
In particular the LHC limits cut away lower mass points and set a
lower limit of $\sim 300 \gev$ for the heavier sleptons in the
Case-R.
This is predominantly due to the LHC limits from slepton pair
production following the decay pattern of \refeq{slepdec} and shown as 
the magenta region in Fig.~\ref{contour_mn1_mslep}. Unlike Case-L, the lighter
masses for both the ``left-'' and ``right-handed'' sleptons
contribute to larger 
production cross-section and lead to the exclusion. Additionally, the
chargino, being mostly wino-like, prefers to decay via light
``left-handed'' sleptons in the kinematically  allowed
region. These points are 
further restricted by LHC searches for chargino production with a
subsequent decay via sleptons as in \refeqs{3ldecviaslep} -
(\ref{chaviaslep}). These limits 
are depicted as red and magenta region in \reffi{contour_mn1_mch}. 
The masses and decay patterns of two sample points from the parameter space
of Case-R scenario is given in \refta{extab3}.
\begin{table}[!htb]
\begin{minipage}[t]{0.35\textwidth}
{\small
\begin{center}
\hspace{-1 cm}
\begin{tabular}[t]{|c||c|c|}
\hline
Sample point                   & 1    & 2   \\
\hline
$\mneu1$                       & 110 & 334\\ 
\hline
$\mneu2$                       & 930 & 632 \\
\hline
$\mneu3$                       & 954  & 1348 \\
\hline
$\mneu4 \sim \mcha2$           & 1101  & 1351 \\
\hline
$\mcha1$                       & 930   & 632 \\
\hline
$m_{\tilde e_1, \tilde \mu_1}$ & 457  & 750   \\
\hline
$m_{\tilde e_2, \tilde \mu_2}$ & 140 & 401 \\
\hline
$m_{\tilde \tau_2}$            & 120 & 349 \\
\hline
$m_{\tilde \tau_1}$            & 462  & 776\\
\hline
$m_{\tilde \nu}$               & 450 &  746 \\
\hline
\end{tabular}
 \end{center}
}
\end{minipage}
\begin{minipage}[t]{0.65\textwidth}
{\small
\begin{center}
\hspace{-1 cm}
\begin{tabular}[t]{|l|c|c||l|c|c|}
\hline
Sample point                           & 1    & 2   & Sample point                   & 1    & 2   \\                                 
\hline
$\br(\neu2 \ra \Sl_1 l$                &22.8  & -          &$\br(\chapm1 \ra \tilde \nu_l l$                      &23.8   &- \\
$\phantom{(\br\neu2}  \ra \Sl_2 l$        &0.4   & -          &$\phantom{\br(\chapm1} \ra \tilde \nu_{\tau_2} \tau$  &18.5   &- \\
$\phantom{\br(\neu2} \ra \Stau2 \tau$  &15.5  &98.8        &$\phantom{\br(\chapm1} \ra \Sl_1 \nu_l$               &18     &-  \\    
$\phantom{\br(\neu2} \ra \Stau1 \tau$  &14.92 & -          &$\phantom{\br(\chapm1} \ra \Stau2 \nu_\tau$           &15     &98.9 \\ 
$\phantom{\br(\neu2} \ra \tilde \nu \nu$ &27.54  &-   &$\phantom{\br(\chapm1} \ra \Stau1 \nu_\tau$           &6.4    &- \\   
$\phantom{\br(\neu2} \ra \neu1 h$             &12.5   & 1.1 &$\phantom{\br(\chapm1} \ra W \neu1)$                  &18.2  &1.1 \\     
$\phantom{\br(\neu2} \ra \neu1 Z)$            &6.2    & 0.1 &                        &       &      \\    

\hline
$\br(\tilde e_2 \ra \neu1 e) $  & 100   & 100  &$\br(\tilde e_1           \ra \neu1 e $  & 100 & 45\\
                                &       &      &$\phantom{\br(\tilde e_1} \ra \neu2 e $) & -   & 18.4\\
                                &       &      &$\phantom{\br(\tilde e_1} \ra \chapm1 \nu_e $) & -   & 36.5     \\
\hline
\end{tabular}
 \end{center}
}
\end{minipage}
\caption{The masses (in $\gev$) and relevant $\br$s (\%) of two sample points
from $\Slpm$-coannihilation Case-R
scenario. Here we show the $\br$s of $\chapm1$ and $\neu2$ to third generation sleptons
separately and that of the first two generations together. Therefore,
$\Sl$ refers to $\tilde e$ and $\tilde \mu$ together. $\nu$ is used to
indicate $\nu_e$, $\nu_\mu$ and $\nu_\tau$ together.
}
\label{extab3}
\end{table}

\begin{figure}[htb!]
	\vspace{2em}
	\centering
	\begin{subfigure}[b]{0.48\linewidth}
	\centering\includegraphics[width=\textwidth]{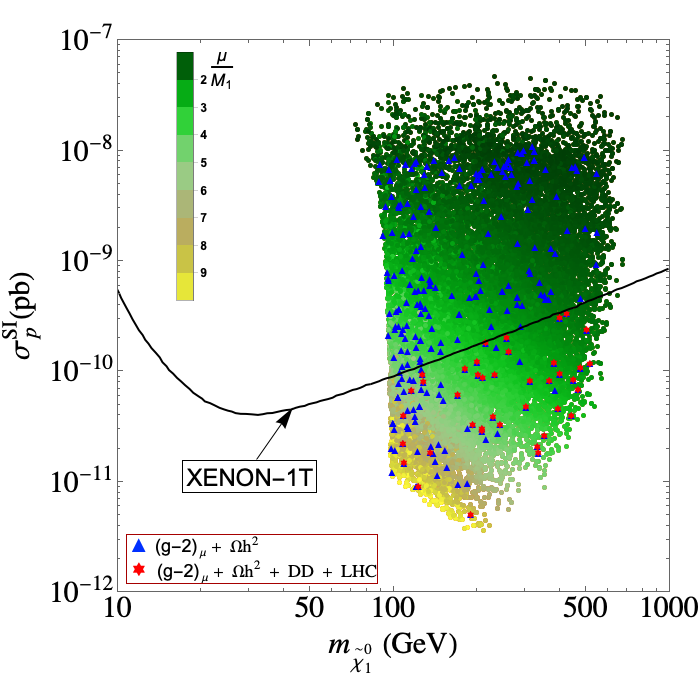}
		\caption{}
		\label{}
	\end{subfigure}
	~
	\begin{subfigure}[b]{0.48\linewidth}
	\centering\includegraphics[width=\textwidth]{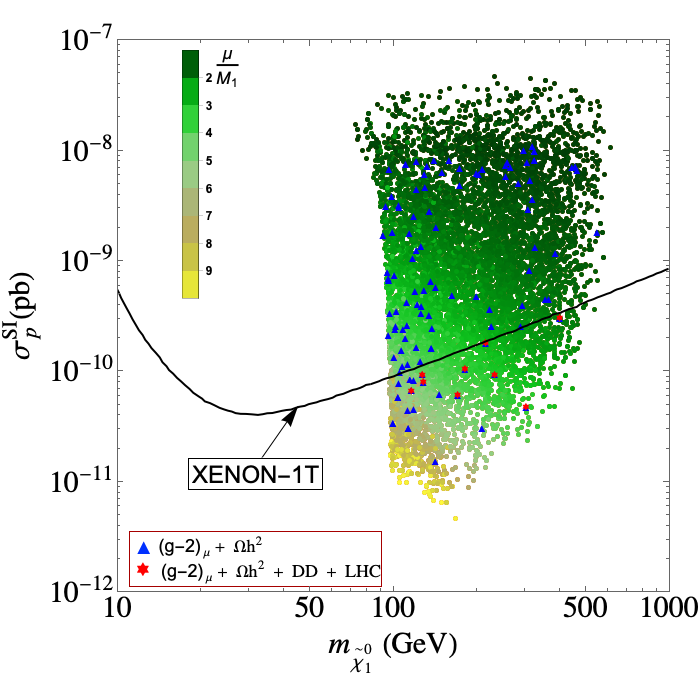}
		\caption{}
		\label{}
	\end{subfigure}
 \caption{The results of our parameter scan in the $\mneu1$-$\ssi$ plane
	for the $\Slpm$-coannihilation 
	Case-R (current (left)
and anticipated future limits (right) from \gmin2).
The color coding as in \protect\reffi{mneu1-ssi-chaco}.
}
\label{mneu1-ssi-caseR}
\end{figure}

The impact of the DD experiments in the Case-R is demonstrated 
in \reffi{mneu1-ssi-caseR}. We show the $\mneu1$-$\ssi$ plane for
current (left) and anticipated future limits (right) from \gmin2. 
The color coding of the points is as in \reffi{mneu1-ssi-chaco}. 
As above, the black line indicates the current DD limits from
XENON1T~\cite{XENON}. 
The general features are similar to the $\Slpm$-coannihilation Case-L scenario:
the scanned parameter space extends from large $\ssi$
values, given for the smallest scanned $\mu/M_1$ values to the
smallest ones, reached for the largest scanned $\mu/M_1$.
However, there is one important change w.r.t.\ Case-L: the \gmin2\
bound tends to drive $\mu$ to lower values, whereas larger values are
preferred by the DD constraint. This ``tension'' results in more intricate
relations among the parameters to be fulfilled in order to meet the
various constraints at once. Although, one
can also see that the relic density constraint is
fulfilled in nearly the whole scanned parameter space.
Given both CDM constraints and the LHC
constraints, the smallest $\mu/M_1$ value we find is smaller than in
the Case-L: 1.7 and 2.0 for the
current and the anticipated future \gmin2\ bound.

\begin{figure}[htb!]
\vspace{2em}
  \centering
  \begin{subfigure}[b]{0.48\linewidth}
    \centering\includegraphics[width=\textwidth]{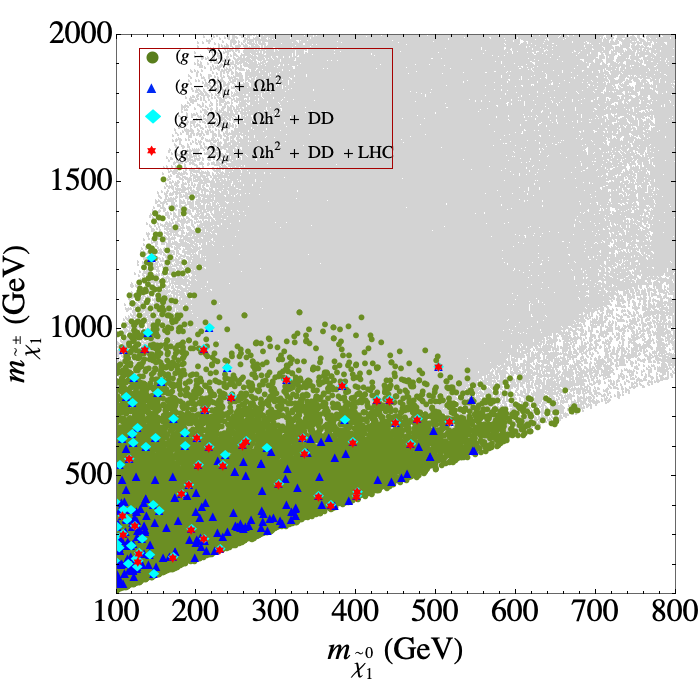}
    \caption{}
    \label{}
  \end{subfigure}
  ~
  \begin{subfigure}[b]{0.48\linewidth}
    \centering\includegraphics[width=\textwidth]{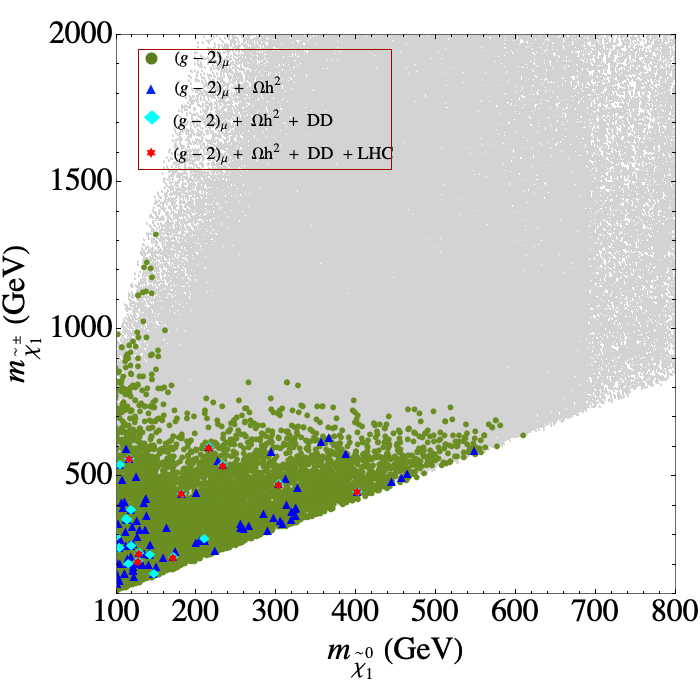}
    \caption{}
    \label{}
  \end{subfigure}
   \caption{The results of our parameter scan in the $\mcha1-\mneu1$
  	plane for the $\Slpm$-coannihilation 
  	Case-R (current (left)
and anticipated future limits (right) from \gmin2).
The color coding as in \protect\reffi{mneu1-msmu2-caseR}.}
  \label{mneu1-mcha1-caseR}
\end{figure}

In \reffi{mneu1-mcha1-caseR} we show the results in the
$\mneu1$-$\mcha1$ plane with the same color coding as
in \reffi{mneu1-msmu2-caseR}. As in the Case-L the \gmin2\ limits on
$\mneu1$ become slightly stronger for larger chargino masses, as expected
from \refeq{amuslep}. The upper limits on the
chargino mass, however, are substantially stronger as in the Case-L. They are
reached at $\sim 1.6 \tev$ for the current and $\sim 1.3 \tev$
for the anticipated future precision in $\amu$. 
On the other hand there are points with very low $\mcha1$, which
are not affected by the LHC searches, which can be understood as follows.
The LHC-excluded points span in principle the entire parameter
space allowed by \gmin2\ constraint with the bounds coming from the searches
discussed in the context of \reffi{mneu1-msmu1-caseR}.
The points with lowest $\mcha1$ values, which are not cut away by
the LHC searches
correspond to the mass hierarchy $\msele1 > \mcha1, \mneu2 > \msele2 > \mstau2$,
with $\msele1$ being relatively large.
Such a configuration implies large
$\br(\chapm1 \to \Stau2 \nu_\tau)$ and
$\br(\neu2 \to \Stau2 \tau)$, leading to a substantially weaker LHC
bounds, as discussed above. For these ``allowed'' points with low $\mcha1$
also $\cha1$-coannihilation contributes relevantly to the CDM relic
density.

\begin{figure}[htb!]
	\vspace{2em}
	\centering
	\begin{subfigure}[b]{0.48\linewidth}
\centering\includegraphics[width=\textwidth]{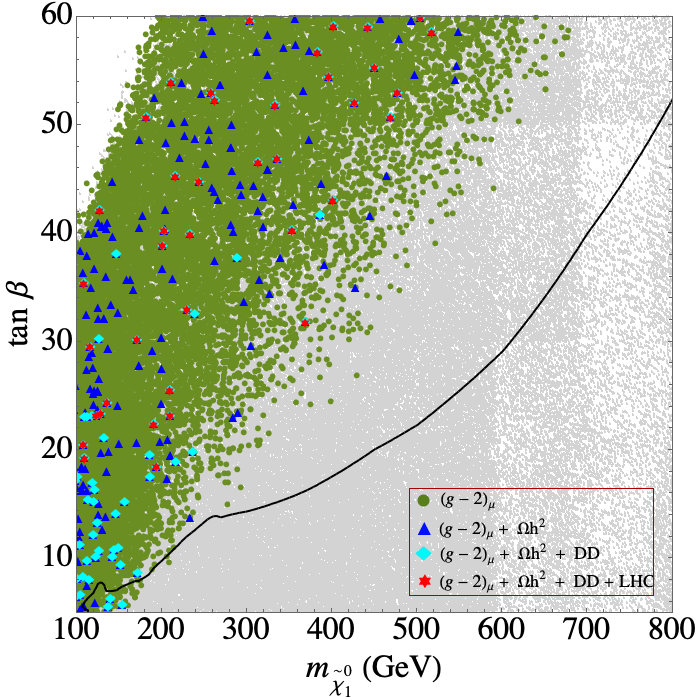}
		\caption{}
		\label{}
	\end{subfigure}
	~
	\begin{subfigure}[b]{0.48\linewidth}
\centering\includegraphics[width=\textwidth]{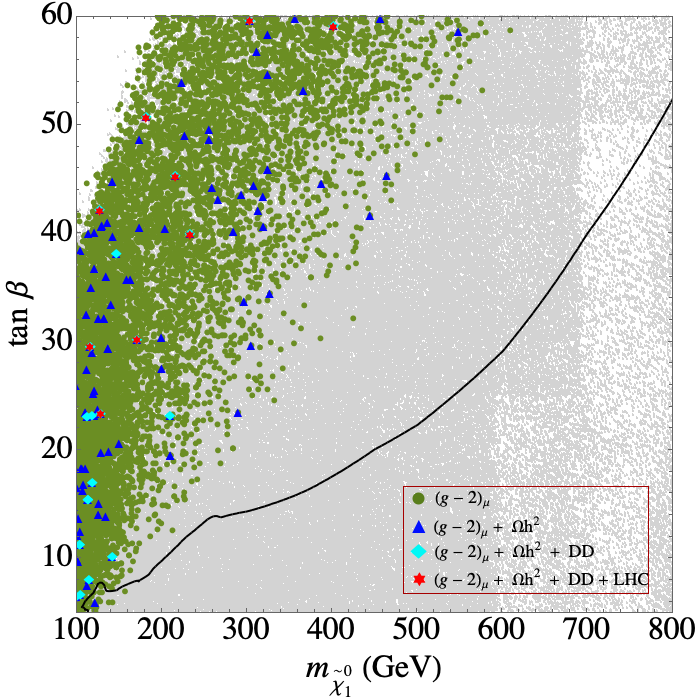}
		\caption{}
		\label{}
	\end{subfigure}
 \caption{The results of our parameter scan in the $\mneu1$-$\tb$
		plane for the $\Slpm$-coannihilation 
	Case-R (current (left)
and anticipated future limits (right) from \gmin2).
The color coding is as in \protect\reffi{mneu1-msmu2-caseR}.
}
\label{mneu1-tb-caseR}
\end{figure}

We finish our analysis of the $\Slpm$-coannihilation Case-R  with the
results in the $\mneu1$-$\tb$ plane, presented in \reffi{mneu1-tb-caseR}. The
overall picture is similar to the previous cases shown
above in \reffis{mneu1-tb-chaco}, \ref{mneu1-tb-caseL}.
Larger LSP masses are allowed for larger $\tb$ values. On the other
hand the combination of small $\mneu1$ and very large $\tb$ values,
$\tb \gsim 40$ leads to stau masses below the LSP mass, which we
exclude for the CDM constraints.
The LHC searches mainly affect parameter points with $\tb \lsim 20$. 
Larger $\tb$ values induce a larger mixing in the third
slepton generation, enhancing the probability for charginos to
decay via staus and thus evading the LHC constraints, see the
discussion in \refse{sec:collider}.
As in \reffi{mneu1-tb-caseL}
we also show the limits from $H/A$ searches at the LHC, where we set
(as above) $\mneu1 = \MA/2$, i.e.\ roughly to the requirement for $A$-pole
annihilation, where points above the black lines are experimentally excluded. 
Comparing Case-R and Case-L, here for the current \gmin2\ limit
substantially less points are passing the current \gmin2\ constraint
below the black line, i.e.\ are potential candidates for $A$-pole
annihilation. The masses reach only up to $\sim 200 \gev$. As in the
two previous cases, see \reffis{mneu1-tb-chaco}, \ref{mneu1-tb-caseL},
these points are reduced in the case of the anticipated future \gmin2\ accuracy
with an upper limit of $\sim 180 \gev$. Together with the already
stronger bounds on $H/A \to \tau\tau$~\cite{Aad:2020zxo} this leaves 
$A$-pole annihilation as a quite remote
possibility with a strict upper bound on $\mneu1$ (see the discussion
in \refse{sec:future-ee}).




\subsection{Lowest and highest mass points}
\label{sec:spectra}

In this section we present some sample spectra for the three cases
discussed in the previous subsections. For each case,
$\cha1$-coannihilation, $\Slpm$-coannihilation Case-L and Case-R, we
present three parameter points that are in agreement with all
constraints (red points): the lowest LSP mass, the highest LSP
with current \gmin2\ constraints, as well as the highest LSP mass
with the anticipated future \gmin2\ constraint.
They will be labeled as "C1, C2, C3", "L1, L2, L3", "R1, R2, R3" for
$\chapm1$-coannihilation, $\Slpm$-coannihilation Case-L and
Case-R,respectively.
While the points are obtained from "random sampling",
nevertheless they give an idea of the mass spectra realized in the
various scenarios.
In particular, the highest mass points give an clear indication on the
upper limits of the NLSP mass.

In \refta{bptab1} we show the $3$ parameter points ("C1, C2,C3") from
$\chapm1$-coannihilation scenario, which are
defined by the six scan parameters: $M_1$, $M_2$, $\mu$, $\tb$ and the
two slepton mass parameters, $\msl{L}$ and $\msl{R}$ (corresponding
roughly to $m_{\tilde e_1, \tilde \mu_1}$ and
$m_{\tilde e_2, \tilde \mu_2}$, respectively).
Together with the
masses and relevant $\br$s we also show the values of the
DM observables and $\amu^{\rm SUSY}$.
Since $\tilde \tau$ is the NLSP in these three cases, the 
contribution from $\tilde \tau$-coannihilation together
with $\chapm1$-coannihilation brings the relic density
to the ballpark value.
For all of the three points, the decays of $\chapm1$ and $\neu2$
to first two generations of sleptons are not kinematically accessible.
Therefore they decay with 100\% BR to third generation charged
sleptons and sneutrinos. This makes them effectively invisible to
the LHC searches looking for 
electrons and muons in the signal. LHC analyses designed to
specifically look for $\tau$-rich final states can prove beneficial
to constrain these points, which are much less powerful, as
discussed above.

\begin{table}[!htb]
{\small
\begin{center}
\hspace{-1 cm}
\begin{tabular}[t]{|c||c|c|c||l||c|c|c|}
\hline
Sample points          & C1   & C2    & C3     &  Sample points                      & C1         & C2   & C3   \\
\hline
\hline
$M_1$                  & 133  & 579 & 430     &$\br(\neu2 \ra \tilde \tau_1 \tau$)     & 100  & 100 &100 \\
\cline{1-4}
$M_2$                  & 144  & 583 & 444     &                                     &      &     & \\  
\cline{1-4}
$\mu$                  & 1329 & 1081  & 1024    &                                     &      &     &  \\
\hline
$\tb$                  &5.1  & 59    & 52.7    &$\br(\chapm1 \ra \tilde \tau_1 \nu_{\tau}$) & 100  & 100 & 100 \\ 
\cline{1-4}
$\msl{L} = \msl{R}$    & 170  &678  & 540   &                                       &      &      &    \\
\cline{1-4}
$\mneu1$               & 129  & 570 & 423     &                                       &      &      &    \\
\cline{1-4}
$\mneu2$               &150   & 605   & 464     &                                      &      &      &    \\
\cline{1-4}
$\mneu3$               &1338  & 1087  & 1032    &                                 &      &      &    \\
\hline
$\mneu4 \sim \mcha2$   & 1341 & 1093  & 1036    &$\br(\Sel1 \ra \neu1 e$             & 20   &  14 & 16.4\\
\cline{1-4}
$\mcha1$               & 150  & 605   & 464     &$\phantom{\br(\Sel1} \ra \neu2 e$                   & 28   &  30 & 28.9   \\
\cline{1-4}
$m_{\tilde e_1,
\tilde \mu_1}$         & 176 & 680   & 542    &$\phantom{\br(\Sel1 } \ra \chapm1 \nu_e$)           & 52   &  55 &54.6  \\
\cline{1-4}
$m_{\tilde e_2,
\tilde \mu_2}$         &176 & 680   & 541 &                                      &      &      &    \\       
\cline{1-4}
$m_{\tilde \tau_1}$    & 140   & 582   & 437   &                                      &      &      &    \\       
\cline{1-4}
$m_{\tilde \tau_2}$    & 205   & 765   & 629    &                                      &      &      &     \\
\hline
$m_{\tilde \nu}$        & 159   & 675 & 536    &$\br(\Sel2 \ra \neu1 e $
& 99.9 & 99.7 & 99.9\\   
\cline{1-4}
$\Omega_{\tilde \chi} h^2$  & 0.118 & 0.121 & 0.118  &$\phantom{\br(\Sel2 } \ra \neu2 e $)                 & 0.1  & 0.3  & 0.1\\
\cline{1-4}
$a_\mu^{\rm SUSY} \tenp{10}$& 21.1  & 15.6  & 20.14   &                                      &      &      &    \\
\cline{1-4}
$\ssi \tenp{10}$            & 0.39  & 2.3  & 1.12    &                                      &      &      &    \\
\hline
\end{tabular}
 \end{center}
}
\caption{The masses (in $\gev$) and relevant $\br$s (\%) of three points
from $\chapm1$-coannihilation
scenario corresponding to the lowest LSP mass, the highest LSP mass
with current \gmin2\ constraints, as well as the highest LSP mass
with the anticipated future \gmin2\ constraint.
Here we show the \br\ of $\chapm1$ and $\neu2$ to third generation sleptons
separately and that of the first two generations together. Therefore,
$\Sl$ refers to $\tilde e$ and $\tilde \mu$ together. $\nu$ is used to
indicate $\nu_e$, $\nu_\mu$ and $\nu_\tau$ together.
Only $\br$s above $0.1$ \% are shown. The values of $\gmin2$ and DM
observables are also shown. $\ssi$ is given in the units of~pb.
}
\label{bptab1}
\end{table}

In \refta{bptab2} we show three parameter points ("L1, L2, L3") taken
from $\Slpm$-coannihilation 
scenario Case-L, defined in the same way as in the $\chapm1$-coannihilation
case. For the point L1, $\chapm1$ and $\neu2$ are higgsino-dominated
with a significant wino component, they are almost mass-degenerate with
$\neu3$. For L2 and L3, on the other hand, $\chapm1$ and $\neu2$ are
wino-like. For this reason, L1 has a 
significant $\br(\neu2 \ra \neu1 h)$ which is absent for L2 and L3.
The large values of $\msmu2$ for the three
points implies that the dominant one-loop contribution to $\gmin2$ comes from
the diagram involving $\chapm1-\Sn$ in the loop.

\begin{table}[!htb]
{\small
\begin{center}
\hspace{-1 cm}
\begin{tabular}[t]{|c||c|c|c||l||c|c|c|}
\hline
Sample points          & L1   & L2   & L3    &  Sample points          & L1         & L2   & L3   \\ 

\hline
\hline
$M_1$                  & 131  & 541 & 508            &$\br(\neu2  \ra \Sl_1 l$      & 32   & 32.4 & 28\\
\cline{1-4}
$M_2$                  & 838  & 793 & 515            &$\phantom{\br(\neu2 } \ra \tilde \tau_1 \tau$ & 17 &18.4 &17.4 \\  
\cline{1-4}
$\mu$                  & 720  & 1365  &  1012        &$\phantom{\br(\neu2 } \ra \tilde \nu \nu$& 34.5 &49.2 &54.6 \\ 
\cline{1-4}
$\tb$                  & 6.95  & 56.7  &56             &$\phantom{\br(\neu2 } \ra \neu1 h$       &  13    & -    & -\\  
\cline{1-4}
$\msl{L}$              & 149 & 548 &509          &$\phantom{\br(\neu2 } \ra \neu1 Z)$       & 3.43     & -    & -\\  
\cline{1-4}
$\msl{R}$               & 1172 & 1278  & 2349        &                                       &      &      &    \\
\cline{1-4}
$\mneu1$               & 126  & 533  & 499           &                                       &      &      &    \\
\cline{1-4}
$\mneu2$               & 706  & 816 &  535           &                                     &      &     & \\  
\cline{1-4}
$\mneu3$               & 731  & 1369 &  1019           &                                     &      &     & \\  
\hline
$\mneu4 \sim \mcha2$   & 889  & 1374 & 1025          &$\br(\chapm1 \ra \tilde \nu_{l_1} l$    & 32 & 33.2 & 39.4\\  
\cline{1-4}
$\mcha1$               & 706 & 816 & 535         &$\phantom{\br(\chapm1 } \ra \tilde \nu_{\tau_1} \tau$& 17 & 17   & 20.4\\  
\cline{1-4}
$m_{\tilde e_1,
\tilde \mu_1}$         & 155  & 549 & 511            &$\phantom{\br(\chapm1 } \ra \Sl_1 \nu_l$             &   23.2  &31.8  & 25.2\\   
\cline{1-4}
$m_{\tilde e_2,
\tilde \mu_2}$         & 1173  & 1279 & 2349           &$\phantom{\br(\chapm1 } \ra \tilde \tau_1 \nu_\tau$&  11.7    &17.7    &15 \\  
\cline{1-4}
$m_{\tilde \tau_1}$    & 155  & 534 &  509           &$\phantom{\br(\chapm1 } \ra W \neu1)$    &   16   &  -   & -\\  
\cline{1-4}
$m_{\tilde \tau_2}$    & 1173 & 1286 & 2350          &                                         &        &      & \\  
\cline{1-4}
$m_{\tilde \nu}$        &135  &544  & 505      &                                     &      &     & \\  
\cline{1-4}
\cline{5-8}
\cline{5-8}
$\Omega_{\tilde \chi} h^2$  & 0.119 & 0.121 & 0.12      &$\br(\Sel1 \ra \neu1 e)$ & 100     & 100    &100 \\  
\cline{1-4}
$a_\mu^{\rm SUSY} \tenp{10}$&19.7 & 14.06 & 21.1       &$\br(\Sel2 \ra \neu1 e $         & 100    & 100    &99.2 \\  
\cline{1-4}
$\ssi \tenp{10}$            & 0.8 & 0.46  & 2.13       &$\phantom{\br(\Sel2 } \ra \neu2 e)$ &   -   & -    & 0.5\\
\hline
\end{tabular}
 \end{center}
}
\caption{The masses (in $\gev$) and relevant $\br$s (\%) of three points
from $\Slpm$-coannihilation
scenario Case-L corresponding to the lowest LSP mass, the highest LSP mass
with current \gmin2\ constraints, as well as the highest LSP mass
with the anticipated future \gmin2\ constraint.
Here we show the \br\ of $\chapm1$ and $\neu2$ to third generation sleptons
separately and that of the first two generations together. Therefore,
$\Sl$ refers to $\tilde e$ and $\tilde \mu$ together. $\nu$ is used to
indicate $\nu_e$, $\nu_\mu$ and $\nu_\tau$ together.
Only $\br$s above $0.1$ \% are shown.
The values of $\gmin2$ and DM
observables are also shown. $\ssi$ is given in the units of~pb.
}
\label{bptab2}
\end{table}

The masses, $\br$s and values of the $\gmin2$ and DM observables
of the three parameter points for the Case-R  ("R1, R2, R3") are shown
in \refta{bptab3}. 
Compared to the points C1 and L1 of the previous two cases, the point
R1 needs a larger value of $\tb$ to satisfy $\gmin2$ constraint. The mass
splitting between $\Smu1$ and $\Smu2$ is also seen to be
smaller than that of Case-L, for reasons discussed in \refse{sec:case2}.
The wino-dominated $\chapm1$ and $\neu2$ preferably decay via
$\Sel1/\Smu1$, which, however, are kinematically forbidden in
these cases. The decays via $\Sel2/\Smu2$, on the other hand, 
are suppressed because of the tiny Yukawa couplings of the first two
generations. Therefore, they decay entirely
to final states involving third generation sleptons, making them
harder to detect.

\begin{table}[!htb]
{\small
\begin{center}
\hspace{-1 cm}
\begin{tabular}[t]{|c||c|c|c||l||c|c|c|}
\hline
Sample points          & R1   & R2    & R3     &  Sample points               & R1         & R2   & R3   \\ 

\hline
\hline
$M_1$                  &111  &525  &408      & $\br(\neu2 \ra \Sl_2 l$            & 0.72 & - & 2.4\\
\cline{1-4}
$M_2$                  &352  &662  &429      &$\phantom{\br(\neu2 } \ra \tilde \tau_2 \tau$ & 93.7   & 96.8 & 97.6\\                   
\cline{1-4}
$\mu$                  &812  &1091  &  822   &$\phantom{\br(\neu2} \ra \neu1 h$              & 4.5  &  2.92   & - \\  
\cline{1-4}
$\tb$                  &20.5 &58.5  & 59     &$\phantom{\br(\neu2} \ra \neu1 Z)$             & 0.99 &  -   & -\\  
\cline{1-4}
$\msl{L}$              &458 & 695   &  794        &      &     &     & \\  
\cline{1-4}
$\msl{R}$              &128 & 591 & 425         &                                       &      &      &    \\
\cline{1-4}
$\mneu1$               &109 & 518  &402     &                                      &      &     &    \\
\hline
$\mneu2$               &367 &685  & 448    & $\br(\chapm1 \ra \Sl_1 \nu_l$             &-     & -     & - \\           
\cline{1-4}
$\mneu3$               &823 & 1098  & 830  &$\phantom{\br(\chapm1} \ra \tilde \tau_2 \nu_{\tau}$ & 94.3  & 97 & 100 \\
\cline{1-4}
$\mneu4 \sim \mcha2$   &828 &1105  & 838   &$\phantom{\br(\chapm1} \ra W \neu1$) & 5.7  & 2.8 &  -\\     
\cline{1-4}
$\mcha1$               &367 &685  &448   &                                      &      &     &    \\
\cline{1-4}
$m_{\tilde e_1,\tilde \mu_1}$&460 &696  &795 &                                      &      &     &    \\
\hline
$m_{\tilde e_2,
\tilde \mu_2}$         &136 & 592 & 428  & $\br(\Sel1 \ra \neu1 e)$             &42      & 95    & 9.2\\                
\cline{1-4}
$m_{\tilde \tau_2}$    &119   &526  &406     &$\phantom{\br(\Sel1} \ra \neu2 e$    &19.6      &  1.7   & 32\\   
\cline{1-4}
$m_{\tilde \tau_1}$    & 464  &747  & 807     &$\phantom{\br(\Sel1} \ra \chapm1 \nu_e)$        &38.3      & 3.2    &58.7 \\  
\cline{1-4}
$m_{\tilde \nu}$        &453 &692  & 792     &                       &      &     &    \\
\cline{1-4}
$\Omega_{\tilde \chi} h^2$  &0.121 &0.121  & 0.121 &                       &      &     &    \\
\hline
$a_\mu^{\rm SUSY} \tenp{10}$&17.5 &14.8  & 17.8    &$\br(\Sel2 \ra \neu1 e)$           &100      &100     & 100\\  
\cline{1-4}
$\ssi \tenp{10}$            &0.23 & 1.2 &  3.1     &                                      &      &     &    \\  
\hline
\end{tabular}
 \end{center}
}
\caption{The masses (in $\gev$) and relevant $\br$s (\%) of three points
from $\Slpm$-coannihilation
scenario Case-R corresponding to the lowest LSP mass, the highest LSP mass
with current \gmin2\ constraints, as well as the highest LSP mass
with the anticipated future \gmin2\ constraint.
Here we show the \br\ of $\chapm1$ and $\neu2$ to third generation sleptons
separately and that of the first two generations together. Therefore,
$\Sl$ refers to $\tilde e$ and $\tilde \mu$ together. $\nu$ is used to
indicate $\nu_e$, $\nu_\mu$ and $\nu_\tau$ together.
Only $\br$s above $0.1$ \% are shown. The values of $\gmin2$ and DM
observables are also shown. $\ssi$ is given in the units of~pb.
}
\label{bptab3}
\end{table}

\clearpage

\section{Prospects for future colliders}
\label{sec:future}

In this section we discuss the prospects of the direct detection of
the (relatively light) EW particles at the approved HL-LHC and at a
hypothetical future $e^+e^-$ collider such as
ILC~\cite{ILC-TDR,LCreport} or CLIC~\cite{CLIC,LCreport}.


\subsection{HL-LHC prospects}
\label{sec:future-pp}

\begin{figure}[htb!]
	\vspace{2em}
	\centering
	\begin{subfigure}[b]{0.48\linewidth}
	\centering\includegraphics[width=\textwidth]{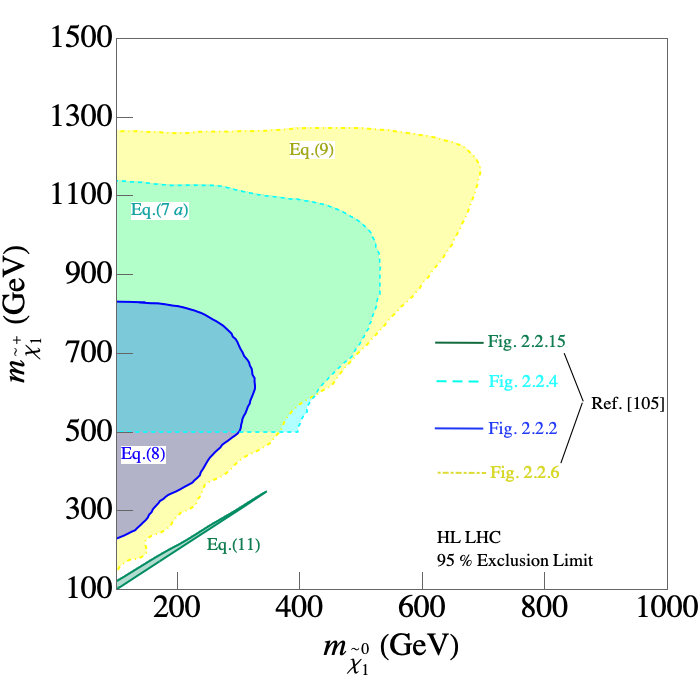}
		\caption{}
		\label{mn1_mc1_exclusion}
	\end{subfigure}
	~
	\begin{subfigure}[b]{0.48\linewidth}
	\centering\includegraphics[width=\textwidth]{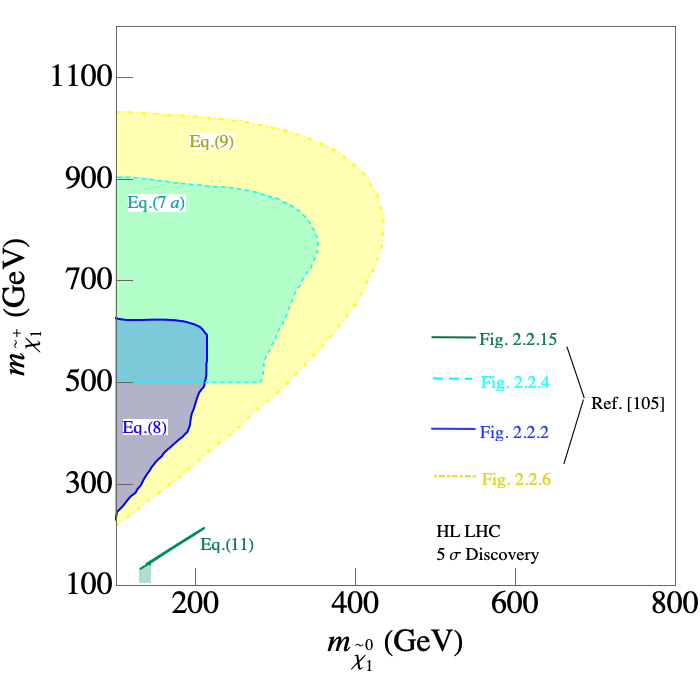}
		\caption{}
		\label{mn1_mc1_discovery}
	\end{subfigure}
\caption{The projected (a) 95\% exclusion limit  and (b) a 5$\sigma$
	discovery reach  
	in the $\mneu1$-$\mcha1$ plane at the HL-LHC.
	The color coding for various decay modes are as
	in \protect\reffi{lhccontours}. 
}
\label{HLLHCcontours}
\end{figure}

The prospects for BSM phenomenology at the HL-LHC have been summarized
in \cite{CidVidal:2018eel} for a 14 TeV run with 3 \iab\ of
integrated luminosity. 
For the direct production of chargino and neutralino through EW
interaction, the projected  95\% exclusion reach as well as a 5$\sigma$
discovery reach have been presented. For the non-compressed scenario, the
electroweak gaugino production via on-shell gauge boson decays has been
analyzed following \refeq{3lviawz} and \refeq{chaviaww}. The
corresponding limits in the $\mneu1$-$\mcha1$ plane is shown in
Fig.~\ref{HLLHCcontours} with cyan and blue shaded regions
respectively. The color coding follows the same convention as in
Fig.~\ref{lhccontours}. The projected exclusion
reach for the cyan region will go twice as high in the chargino mass
range, covering values
as high as 1~TeV for $\mneu1 \lesssim 500 \gev$. However, the most
significant improvement can be observed in 
the gaugino productions via on-shell $W$ and Higgs decays as
in \refeq{chaviawh}, shown as yellow shaded area. On the other hand, the
same search channel has a projected $5\,\sig$ reach up to
$\mcha1 \lesssim 1 \tev$ and $\mneu1 \lesssim 500 \gev$. Therefore,
this search channel with an on-shell Higgs decaying to $b\bar b$
combined with the future \gmin2\ accuracy and DM constraints
can conclusively probe ``almost'' the entire allowed parameter region
of $\Slpm$-coannihilation Case-R scenario
and a significant part of the same parameter space
for Case-L scenario (see
Fig.~\ref{mneu1-mcha1-caseL} and Fig.~\ref{mneu1-mcha1-caseR}) at the
HL-LHC. 
We note that no update was provided in \cite{CidVidal:2018eel}
for the most relevant search channels, the decay of
charginos/neutralinos via intermediate sleptons with $3l$ and $2l$ final
states, see \refeqs{3ldecviaslep}, (\ref{chaviaslep}). A clearer picture
of the HL-LHC prospects for the physics case under investigation could
be drawn with an experimental analysis of these channels.

Similarly, no prospects for the scalar electron or muon
production at the HL-LHC have been reported yet. However, the analysis
for compressed higgsino-like spectra may exclude
$\mneu2 \sim \mcha1 \sim 350 \gev$ with mass gap as low as 2~GeV for
$\mcha1$ around 100~GeV following the decay pattern
of \refeq{offshellwz}. Hence, a substantial parameter region can be
curbed for the $\cha1$-coannihilation case (see
\reffi{mn1_mc1_chaco}) in the absence of a signal in the compressed
scenario analysis with soft leptons at the final state.


\subsection{ILC/CLIC prospects}
\label{sec:future-ee}

Direct production of EW particles at $e^+e^-$ colliders clearly
profits from a higher center-of-mass energy, $\sqrt{s}$. Consequently,
we focus here on the two proposals for linear $e^+e^-$ colliders,
ILC~\cite{ILC-TDR,LCreport} and CLIC~\cite{CLIC,LCreport}, which can
reach energies up to $1 \tev$, and $3 \tev$, respectively.
We evaluate the cross-sections for the various SUSY pair
production modes for the energies currently foreseen in the run plans
of the two colliders.  The anticipated energies and integrated
luminosities are listed in \refta{tab:ee-sqrtS}. The cross-section
predictions are based on tree-level results, obtained as
in~\cite{Heinemeyer:2017izw,Heinemeyer:2018szj}, where it was shown
that the full one-loop corrections can amount up to 10-20\%\,%
\footnote{Including full one-loop corrections here as 
in \cite{Heinemeyer:2017izw,Heinemeyer:2018szj} would have required to
determine the preferred choice of the renormalization scheme for each
point individually (see \cite{Fritzsche:2013fta} for details), which
goes beyond the scope of this analysis.}%
. We do not attempt any rigorous experimental analysis,
but follow the idea that to a good approximation
final states with the sum of the masses smaller than the
center-of-mass energy can be
detected~\cite{Berggren:2013vna,PardodeVera:2020zlr,Berggren:2020tle}. 
We also note that in case of several EW SUSY particles in reach of an
$e^+e^-$ collider, large parts of the overall SUSY spectrum can be
measured and fitted~\cite{Baer:2019gvu}.
\begin{table}[!htb]
\begin{center}
\renewcommand{\arraystretch}{1.4}
\begin{tabular}{|c|c|c||c|c|c|}
\hline
Collider & $\sqrt{s}$ [GeV] & $\cL_{\rm int}$ $[\iab]$ & 
Collider & $\sqrt{s}$ [GeV] & $\cL_{\rm int}$ $[\iab]$ \\
\hline
ILC & 250 & 2 & CLIC & 380 & 1 \\
    & 350 & 0.2 &    & 1500 & 2.5 \\
    & 500 & 4 &      & 3000 & 5 \\
    & 1000 & 8 &     &      & \\
\hline
\end{tabular}
\caption{Anticipated center-of-mass energies, $\sqrt{s}$ and
    corresponding integrated luminosities, $\cL_{\rm int}$ at
    ILC~\cite{Barklow:2015tja,Fujii:2017vwa} and
    CLIC~\cite{Robson:2018zje} (as used in \cite{deBlas:2019rxi}).}
\label{tab:ee-sqrtS}
\renewcommand{\arraystretch}{1.2}
\end{center}
\end{table}

In the following we show the results for certain EW-SUSY production
cross-sections for a fixed $\sqrt{s}$ (according
to \refta{tab:ee-sqrtS}). We do not show the production cross-sections
for the mono-photon signal
$e^+e^- \to \neu1\neu1 (+\ga)$ (where the ISR photon is required to
detect the invisible final state). From the upper limits on $\mneu1$ as
obtained in \refse{sec:results} (about $570 \gev$ 
for $\cha1$-coannihilation, about $540 \gev$ for
$\Slpm$-coannihilation Case-L and about
$520 \gev$ in Case-R, for the current \gmin2\ constraint), it can be
inferred that with $\sqrt{s} = 1000 \gev$
a considerable part of the spectrum can be
covered. The reach becomes even stronger in the case of the future
anticipated \gmin2\ constraint, where the upper limits on $\mneu1$ go
down to $\sim 430, 500, 410 \gev$, respectively, covering effectively
the full spectrum at a $1000 \gev$ collider.

\begin{figure}[htb!]
  \centering
  \begin{subfigure}[b]{0.48\linewidth}
    \centering\includegraphics[width=\textwidth]{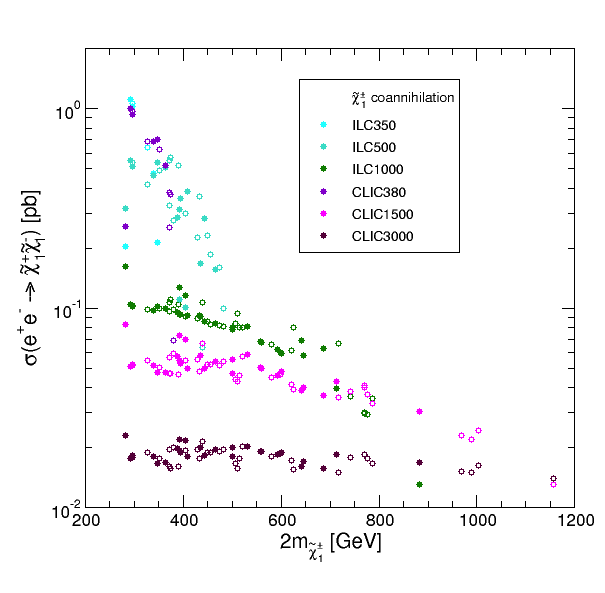}
    \caption{}
    \label{}
  \end{subfigure}
  ~
  \begin{subfigure}[b]{0.48\linewidth}
    \centering\includegraphics[width=\textwidth]{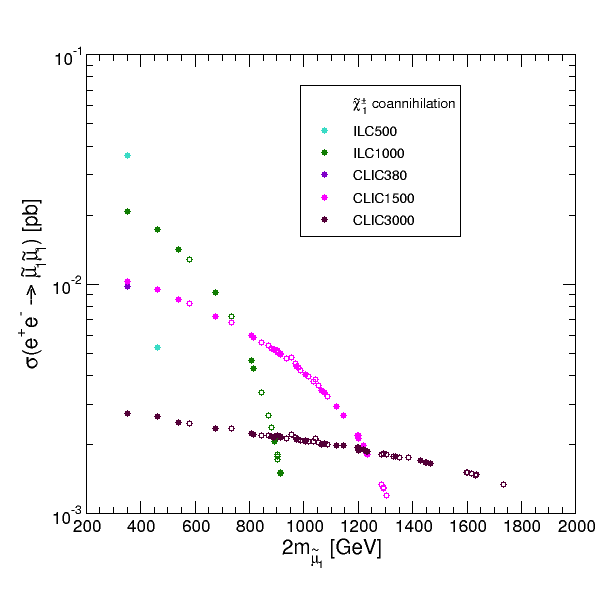}
    \caption{}
    \label{}
  \end{subfigure}
  \caption{cross-section predictions for
  $e^+e^- \to \chapm1\champ1$ (left) and
  $e^+e^- \to \Smu1\Smu1$ (right) in the $\cha1$-coannihilation case at
  the ILC and CLIC as a function of the sum of the final state masses.
  Open (filled) circles indicate agreement with \gmin2\ at the current
  (anticipated future) accuracy.}
\label{ee-char}
\end{figure}

We start the numerical investigations in \reffi{ee-char} with the
cross-section predictions for
$e^+e^- \to \chapm1\champ1$ (left) and
$e^+e^- \to \Smu1\Smu1$ (right) in the $\cha1$-coannihilation case as a
function of the sum of the final state masses. The different shades of
green (violet) indicate the cross-sections at the various ILC (CLIC)
energies. Open (filled) circles indicate agreement with \gmin2\ at the current
(anticipated future) accuracy. It can be observed that the lower-energy
stages of the ILC and CLIC, $\sqrt{s} \le 500 \gev$, cover only a very
small part of the predicted mass spectrum. ($\sqrt{s} = 250 \gev$ does
not yield any accessible parameter point in our analysis.) Higher
energies, on the other hand, as can be reached in principle at future
$e^+e^-$ colliders cover part or even the full predicted spectrum. 
This holds particularly for the CLIC energies for the parameter points
assuming the
current \gmin2\ constraint. On the other hand, even the ILC1000 can
cover the full predicted $\mcha1$~spectrum in the case of the future
anticipated \gmin2\ constraint. All obtained cross-section predictions
for the kinematically accessible parameter points are above $10^{-2}$~pb
for chargino production and above $10^{-3}$~pb for smuon pair
production. For each \iab\ of integrated luminosity this corresponds to
10000 (1000) events for chargino (smuon) pair production, which should
make these particles easily accessible, see \refta{tab:ee-sqrtS}, if
they are in the kinematic reach of the collider.

\begin{figure}[htb!]
  \centering
  \begin{subfigure}[b]{0.48\linewidth}
    \centering\includegraphics[width=\textwidth]{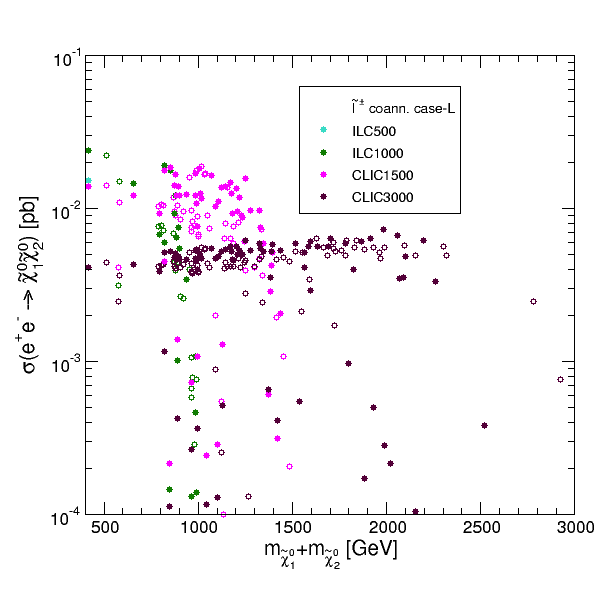}
    \caption{}
    \label{}
  \end{subfigure}
  ~
  \begin{subfigure}[b]{0.48\linewidth}
    \centering\includegraphics[width=\textwidth]{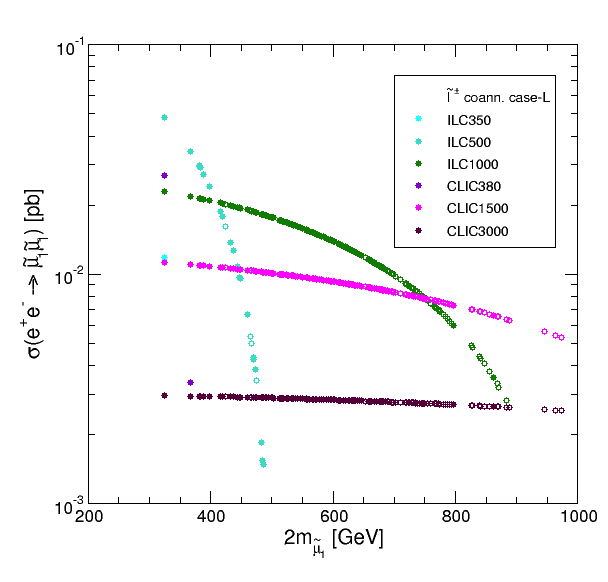}
    \caption{}
    \label{}
  \end{subfigure}
  \caption{cross-section predictions for
  $e^+e^- \to \neu1\neu2$ (left) and
  $e^+e^- \to \Smu1\Smu1$ (right) in the $\Slpm$-coannihilation Case-L at
  the ILC and CLIC as a function of the sum of the final state masses.
  Symbols are as in \protect\reffi{ee-char}.}
\label{ee-slep-L}
\end{figure}

The situation is similar, but somewhat less encouraging for
$\Slpm$-coannihilation Case-L, as presented in \reffi{ee-slep-L}. The left
(right) plot shows the cross-section predictions for
$e^+e^- \to \neu1\neu2$ and $e^+e^- \to \Smu1\Smu1$ with the same
symbol/color coding as in \reffi{ee-char}. Here highest CLIC energies
are needed to cover the full predicted spectrum. On the other hand, the
lighter sleptons are closer in mass to the LSP in this scenario.
Consequently, even the ILC500 can cover a substantial part of the
predicted $\msmu1$ spectrum, and $\sqrt{s} \lsim 1000 \gev$ is
sufficient to cover all predicted mass values.

\begin{figure}[htb!]
  \centering
  \begin{subfigure}[b]{0.48\linewidth}
    \centering\includegraphics[width=\textwidth]{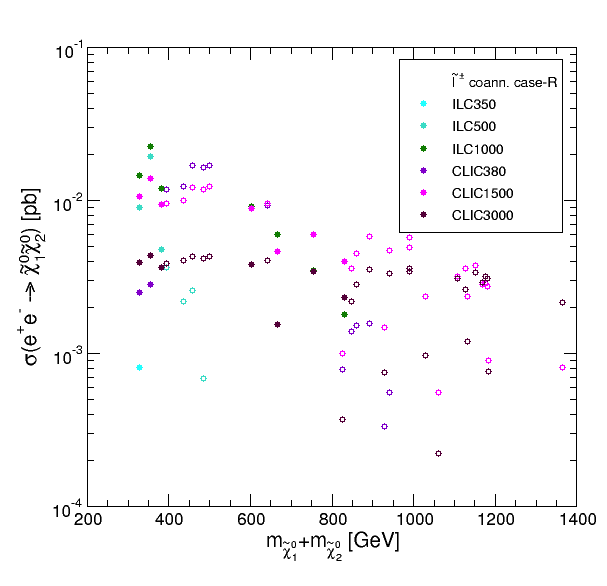}
    \caption{}
    \label{}
  \end{subfigure}
  ~
  \begin{subfigure}[b]{0.48\linewidth}
    \centering\includegraphics[width=\textwidth]{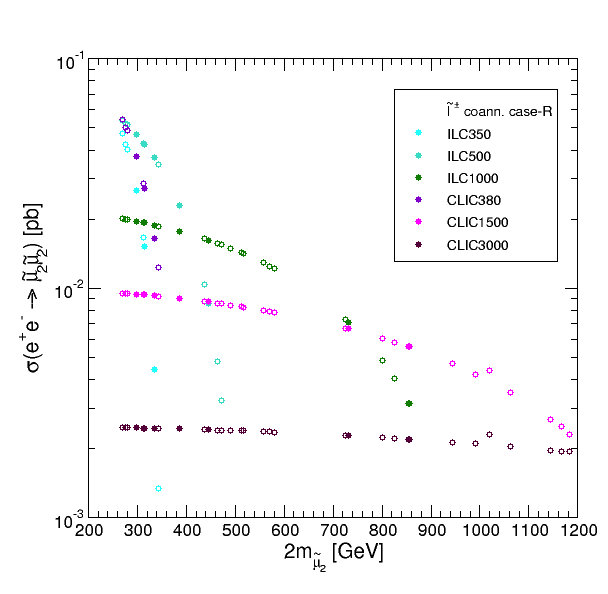}
    \caption{}
    \label{}
  \end{subfigure}
  \caption{cross-section predictions for
  $e^+e^- \to \neu1\neu2$ (left) and
  $e^+e^- \to \Smu2\Smu2$ (right) in the $\Slpm$-coannihilation Case-R at
  the ILC and CLIC as a function of the sum of the final state masses.
  Symbols are as in \protect\reffi{ee-char}.}
\label{ee-slep-R}
\end{figure}

Finally in \reffi{ee-slep-R} we show the $\Slpm$-coannihilation
Case-R. As argued in the previous subsection, in this case some EW SUSY
particles are driven to relatively low mass values. This is visible
in \reffi{ee-slep-R}, where the left
(right) plot shows the cross-section predictions for
$e^+e^- \to \neu1\neu2$ and $e^+e^- \to \Smu2\Smu2$ with the same
symbol/color coding as in \reffi{ee-char}. In the case of $\neu1\neu2$
production even with the current \gmin2\ constraint a collider with
$\sqrt{s} = 1500 \gev$ is sufficient to cover the full spectrum. With
the anticipated future \gmin2\ accuracy only mass sum values $\lsim 800 \gev$
are reached, and even the ILC500 can cover a relevant part of the
spectrum. Most cross-sections are at or above $10^{-3}$~pb, i.e.\ at least
1000 events per \iab\ would be produced.
The situation is similar for $\Smu2\Smu2$ production, although with even
higher production cross-sections. The ILC1000 could cover the full
spectrum corresponding to the anticipated future \gmin2\ accuracy, and
some part of the spectrum would be accessible even at the ILC350 (or
circular $e^+e^-$ colliders reaching this energy).

\smallskip
The above shown example cross-sections clearly show the anticipated
future accuracy on \gmin2\ has the power to sharpen the {\it upper}
limits on EW SUSY 
particles sufficiently, such that at least some particles are guaranteed
to be discovered at the higher-energy stages of the ILC and/or CLIC.
(The LSP should be accessible in almost any case even with the current \gmin2\
accuracy.) If
the future \gmin2\ constraint confirms the deviation of $\amu^{\rm exp}$
from the SM prediction, the case for future $e^+e^-$ colliders is
clearly strengthened.

\medskip
As discussed in \refse{sec:scan} we have not considered the
possibility of $Z$~or $h$~pole annihilation to find agreement of the
relic DM density with the other experimental measurements. 
However, it should be noted that in this context an LSP with
$M_1 \sim \mneu1 \sim \MZ/2$ or $\sim \Mh/2$ would yield a detectable
cross-section $e^+e^- \to \neu1\neu1\ga$ in any future high-energy $e^+e^-$
collider. Furthermore, depending on the values of $M_2$ and $\mu$,
this scenario likely yields other clearly detectable EW-SUSY
production cross-sections at future $e^+e^-$ colliders. We leave this
possibility for future studies.

On the other hand, the possibility of $A$-pole annihilation was
discussed for all three scenarios. While it appears a rather remote
possibility, it cannot be fully excluded by our analysis. However,
even in the ``worst'' case of $\Slpm$-coannihilation Case-L an upper
limit on $\mneu1$ of $\sim 260 \gev$ can be set. While not as low as
in the case of $Z$~or $h$-pole annihilation, this would still offer
good prospects for future $e^+e^-$ colliders. We leave also this
possibility for future studies.


\section {Conclusions}
\label{sec:conclusion}

The electroweak (EW) sector of the MSSM, consisting of charginos,
neutralinos and scalar leptons can account for a variety of experimental
data. Concerning the CDM relic abundance, the MSSM
offers a natural candidate, the lightest neutralino,~$\neu1$,
while satisfying the bounds from
DD experiments which have yielded negative results so far.
As a result of comparatively small
production cross-sections, a relatively light EW sector of the
MSSM is also in agreement with
the latest experimental exclusion limits from the LHC. Most importantly, the EW sector of the
MSSM can account for the long-standing $3-4\,\sig$ discrepancy of \gmin2.
Improved experimental results are expected in the course of 2020 by the 
publication of the Run~1 data of the ``MUON G-2'' experiment.

In this paper, under the assumption that the $\neu1$
provides the full DM relic abundance we first analyzed which mass ranges of
neutralinos, charginos and sleptons are in agreement with all
relevant experimental data: the current limit for \gmin2\,, the relic
density bounds, the DD experimental bounds, as well as the LHC searches
for EW SUSY particles. Concerning the latter we included all relevant
existing data, mostly relying on re-casting via \CM, where several
channels had to be newly implemented. 
We analyzed three scenarios, depending on the mechanism that brings the
relic density in agreement with the experimental data:
$\cha1$-coannihilation, $\Slpm$-coannihilation with the mass of the
``left-handed'' (``right-handed'') slepton close to $\mneu1$, Case-L
(Case-R). We find in all three cases a clear upper limit on $\mneu1$.
While for $\cha1$-coannihilation this is $\sim 570 \gev$, for
$\Slpm$-coannihilation Case-L $\sim 540 \gev$ and for Case-R
values up to $\sim 520 \gev$ are allowed. Similarly,
upper limits to masses of the coannihilating SUSY particles are found as,
$\mcha1 \lsim 610 \gev$, 
$\msl{L} \lsim 550 \gev$, 
$\msl{R} \lsim 590 \gev$. For the latter, in the
$\Slpm$-coannihilation case-R, the upper limit on the
lighter $\stau$ is even lower, $\mstau2 \lsim 530 \gev$.
The current \gmin2\ constraint also yields limits on the rest of the EW
spectrum, although much loser bounds are found. As an example, for
$\cha1$-coannihilation we find $\msl{L} \lsim 900 \gev$, for
$\Slpm$-coannihilation Case-L $\mcha1 \lsim 3 \tev$ and
$\mcha1 \lsim 1 \tev$ for Case-R. These upper bounds set clear
collider targets for the HL-LHC and future $e^+e^-$ colliders.

In a second step we assumed that the new result of the
Run~1 of the ``MUON G-2'' collaboration at Fermilab yields a precision
comparable to the existing experimental result with the same central
value. We analyzed the potential impact of the combination of the Run~1
data with the existing result on the allowed MSSM parameter space.  
We find that the upper limits on the LSP mass are decreased
to about $425 \gev$ for 
$\chapm1$-coannihilation, $500 \gev$ for $\Slpm$-coannihilation Case-L
and $400 \gev$ in Case-R,
sharpening the collider targets substantially.
Similarly, the upper limits on the NLSP masses go down to about
$470 \gev$, $510 \gev$ and $410 \gev$ in the three cases that we have explored.

For the HL-LHC we have briefly reviewed the anticipated future upper
limits and $5\,\sig$ discovery regions. We find that 
the chargino/neutralino production at the HL-LHC via on-shell~$W$ and
Higgs decays have a substantial impact on the allowed parameter space of
$\Slpm$-coannihilation scenarios. 
In particular, the  fate of Case-R can be conclusively determined with
only this search channel. 
On the other hand, the compressed spectrum searches for
chargino-neutralino production may become 
important to probe the parameter space of $\chapm1$-coannihilation region.

Concerning future high(er) energy $e^+e^-$ colliders, we have evaluated
the production cross-sections for the anticipated center-of-mass
energies of ILC and CLIC.
The LSP should be accessible in almost any case at $\sqrt{s} = 1000 \gev$
even with the current \gmin2\ accuracy and can effectively be fully
covered with the anticipated future accuracy. 
Moreover, since the anticipated
future accuracy on \gmin2\ has the potential to narrow down the {\it upper}
limits on EW SUSY particles sufficiently,
the analyzed example cross-sections show
that at least some particles are guaranteed
to be discovered at the higher-energy stages of the ILC and/or CLIC.
Therefore, if the future \gmin2\ constraints confirms the deviation
of $\amu^{\rm exp}$ from the SM prediction, our findings strongly motivate
the need of future $e^+e^-$ colliders.

\medskip
While we have attempted to cover nearly the full set of possibilities
that the EW spectrum of the MSSM can fulfill all the various
experimental constraints, our
studies can be extended/completed in the following ways. One can analyze
the cases of:
(i) complex parameters in the chargino/neutralino sector (then also
taking EDM constraints into account);
(ii) different soft SUSY-breaking parameters in the three generations of
sleptons, and/or between the left- and right-handed entries in the case
of $\chapm1$-coannihilation;
(iii) $A$-pole annihilation, in particular in the case of
$\Slpm$-coannihilation for very low $\mneu1$ and $\tb$ values;
(iv) $h$-~and $Z$-pole annihilation, which could be realized for
sufficiently heavy sleptons;
(v) requiring the CDM constraint only as upper limit.
We leave these analyses for future work.

\medskip
In this paper we have analyzed in particular the impact of \gmin2\
measurements on the EW SUSY spectrum. While the current experimental and
theoretical situation is clear, the upcoming measurements of the
``MUON G-2'' collaboration were shown to have the strong potential of
sharpening the future collider experiment prospects. All this hinges, of
course, on the central value the collaboration will observe (where we
used the simplest assumption of the same central value as in the current
experimental data). We are eagerly awaiting the new ``MUON G-2'' result
to illuminate further the possibility of relatively light EW BSM
particles.


\subsection*{Acknowledgments}

We thank
M.~D'Onofrio, 
J.~List
and
D.~St\"ockinger
for helpful discussions.
We thank J.~S.~Kim for help with the implementation of new LHC search
channels to \CM.
We thank C.~Schappacher for the calculation of the $e^+e^-$ EW-SUSY
production cross-sections.
I.S.\ gratefully thanks S.~Matsumoto for the cluster facility.
I.S.\ acknowledges the warm hospitality of the
IFT, Madrid and thanks the organizers of the workshop ``Opportunities
at Future High Energy Colliders'' for the invitation where this work
was initiated. The work of I.S.\ is supported by World Premier
International Research Center Initiative (WPI), MEXT, Japan. 
The work of S.H.\ is supported in part by the
Spanish Agencia Estatal de Investigaci{\' o}n (AEI) and the EU Fondo Europeo de
Desarrollo Regional (FEDER) through the project FPA2016-78645-P and in part by
the “Spanish Red Consolider MultiDark” FPA2017-90566-REDC. 
The work of M.C.\ and S.H.\ is supported in 
part by the MEINCOP Spain under contract FPA2016-78022-P and in part by
the AEI through the grant IFT Centro de Excelencia Severo Ochoa SEV-2016-0597.
  

\newcommand\jnl[1]{\textit{\frenchspacing #1}}
\newcommand\vol[1]{\textbf{#1}}

\newpage{\pagestyle{empty}\cleardoublepage}


\end{document}